# Host Cell Factors Necessary for Influenza A Infection: Meta-Analysis of Genome Wide Studies

**Juliana S. Capitanio and Richard W. Wozniak**
Department of Cell Biology, Faculty of Medicine and Dentistry, University of Alberta

**Abstract:** The Influenza A virus belongs to the Orthomyxoviridae family. Influenza virus infection occurs yearly in all countries of the world. It usually kills between 250,000 and 500,000 people and causes severe illness in millions more. Over the last century alone we have seen 3 global influenza pandemics. The great human and financial cost of this disease has made it the second most studied virus today, behind HIV.
Recently, several genome-wide RNA interference studies have focused on identifying host molecules that participate in Influenza infection. We used nine of these studies for this meta-analysis. Even though the overlap among genes identified in multiple screens was small, network analysis indicates that similar protein complexes and biological functions of the host were present. As a result, several host gene complexes important for the Influenza virus life cycle were identified. The biological function and the relevance of each identified protein complex in the Influenza virus life cycle is further detailed in this paper.

## Background

### Influenza virus

Viruses are the simplest life form on earth. They parasite host organisms and subvert the host cellular machinery for different steps of their life cycle. One such example is the Influenza A virus, belonging to the Orthomyxoviridae family. It has a negative-sense, single-stranded, and segmented RNA genome stored in a ribonucleoprotein complex within the viral core. This complex contains the viral polymerases PB1, PB2 and PA bound to the viral genome via nucleoprotein (NP). The viral core is enveloped by a lipid membrane derived from the host cell. The viral protein M1 underlies the membrane and anchors NEP/NS2. Hemagglutinin (HA), neuraminidase (NA), and M2 proteins are inserted into the envelope, facing the viral exterior.

A diagram of the Influenza A virus life cycle from the Reactome database (1) can be seen in figure 1. To initiate infection and replication the Influenza virus binds to molecules of sialic acids on the surface of a host cell. Binding at the cell surface, the virus is internalized by receptor-mediated endocytosis, most commonly clathrin mediated endocytosis. The low pH in the endosome promotes the fusion of the viral and endosomal membranes, releasing the viral ribonucleoprotein complexes (vRNPs) into the cytoplasm. The vRNPs, containing the viral genome coated by NPs and associated to the viral polymerases must be imported into the nucleus for transcription and replication. Nucleoprotein has been shown to possess a nuclear localization signal that is sufficient to activate the nuclear import of the vRNPs. Once in the nucleus the viral RNA will be replicated into new vRNA through a positive-sense intermediate (cRNA) and transcribed into viral messenger RNAs. The viral mRNAs can be spliced into different transcripts and they are exported from the nucleus and translated into viral proteins in the cytoplasm. Some of the viral proteins re-enter the nucleus to assemble vRNPs. Others are directed to the cell membrane through the Golgi apparatus where they form lipid rafts. The vRNPs are then also directed to the same region of the cell membrane where the budding of new viral particles occurs.

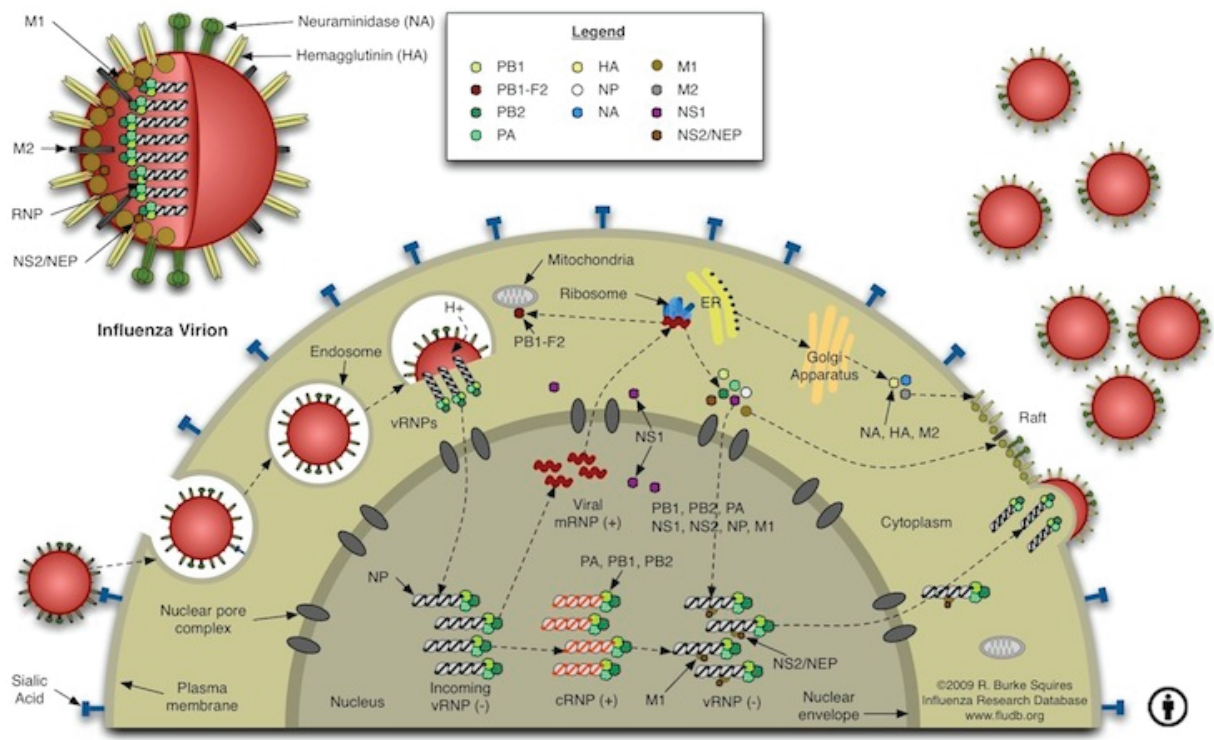

**Figure 1:** **Influenza virus life cycle.** Illustration of the influenza virus replication cycle from the Reactome database (1).

All the steps described above significantly depend on the host cell machinery, therefore characterizing the involvement of host factors in infections is of great scientific interest. It can better clarify the viral life cycle and allow the identification of important drug targets to treat the diseases caused by these infections.

Influenza virus infection occurs yearly in all countries of the world. It usually kills between 250,000 and 500,000 people and causes severe illness in millions more. Over the last century alone we have seen 3 global influenza pandemics. The great human and financial cost of this disease has made it the second most studied virus today, behind HIV (2).

### Genome-wide Influenza dependency host factor screens

In order to gain new insight into the life cycle of Influenza viruses and to identify potential therapeutic targets for infection, several laboratories have focused in genome-wide screen strategies. Screens based on four different strategies were compiled for this meta-analysis (table I).

### Genome-wide RNAi screening

Most of the data collected was based on genome-wide RNAi experiments, this type of screen employs systematic knockdown of known genes and evaluates the effect depletion of that given target has in infection.

The first Influenza infection RNAi screen was reported by Hao, et al in 2008 (3). This study was based on infection of Drosophila cells; it used a modified A/WSN/33 influenza strained capable of infecting Drosophila cells and producing a fluorescent reporter in infected cells (table I). Out of the 13,071 genes targeted 175 were identified as affecting reporter expression and therefore Influenza infection. A list of 130 human homologues to the 175 genes identified was used in this meta-analysis (table SI). Among these are ATP6V0D1, COX6A1 and NFX1, genes whose importance in Influenza infection was confirmed in mammalian cell systems (3).

Using genome wide RNAi screens targeting over 17,000 genes in mammalian cells other groups also investigated the importance of host factors for Influenza infection. Konig, et al (4) and Karlas, et al (5) both used the human A549 lung cell line infected with A/WSN/33 Influenza strain for their screens. Brass, et al (6) made use of an osteosarcoma cell line infected with A/PR8/34 strain of Influenza (table I). As seen in the experiments performed by Hao, et al (3) the as-say readout for two of these screens was quantification of a virally encoded reporter protein (4) or a viral protein (6). The three screens described above (Hao, et al (3); Konig, et al (4) and Brass, et al (6)) can evaluate only part of the viral life cycle, up to the point of viral protein translation. They do not assess factors needed for viral budding or the infectivity of the produced viruses. Circumventing this problem Karlas, et al (5) evaluated the entire viral life cycle by quantifying the infectivity of viruses produced by the RNAi containing cells. One last genome-wide RNAi screen used in this meta-analysis came from the work of Shapira, et al (7). In a different approach, this group started out by using a combination of yeast two hybrid, to detect host-viral protein interactions, and microarray, to evaluate how infection altered the expression of host genes. These identified host genes affected by Influenza viruses were then targeted by RNAi to determine their importance in infection. This screen also assessed the infectivity of viruses produced by RNAi treated cells, making it possible to evaluate the entire viral life cycle.

These screens identified hundreds of novel host factors required for Influenza infection, they are summarized in table I and table SI. Only the final results presented in each paper, after application of all statistical tests, were used in this meta-analysis (table SI).

### Random homozygous gene perturbation (RHGP)

In a different perspective on the issue, Sui, et al (8) looked to identify host genes that contributed to resistance to Influenza infection. They used RHGP to generate a library of randomly silenced or overexpressed genes in MDCK cells. This library was infected with a high dose of Influenza virus (A/Udorn/72 strain) that would lead to cell death. Surviving cells presented altered expression of a gene that rendered them resistant to Influenza infection, allowing the identification of 110 such host cell factors (table I and table SI).

### Viral particles proteomics

All the proteins controlling viral exit, entrance and initial replication in the host cell are thought to be contained within the viral particle. Two reports by Shaw, et al (9) and Song, et al (10) demonstrate that these viral particles contain not only viral proteins, but also several host proteins. In order to identify these host proteins packaged within the Influenza virion these two groups purified Influenza virus particles (9) or recombinant Influenza virus-like particles (10) and used mass spectrometry to identify all proteins present. Over 30 host proteins were identified in each screen, with the presence of cytoskeletal proteins been a common theme on both (table I and table SI). Identification of host proteins specifically incorporated into virions can indicate their requirement for infection, making them good targets for antiviral drugs.

### Host-pathogen protein-protein interaction database (HPIDB)

Host-pathogen protein interactions have a very important role in infection. From infection initiation to the budding of new viral particles the virally encoded proteins interact with several host factors. These interactions have been described extensively in the literature and in numerous databases. The HPIDB integrates these experimental results into a single, non-redundant resource (11). It collects data on several pathogens, for this analysis all information available for Influenza virus (A/PR/8/34) was downloaded from their server (12) and identified genes can be seen in table SI.

**Table I: Genome-wide data used in this meta-analysis.** Description of Influenza virus host dependency factor screens and their main results.

| Authors | Cell line | Viral Strain | Identified host factors | Screen type | Main affected proteins / pathways |
|---|---|---|---|---|---|
| Brass et al, 2009 | U2OS | A/PR/8/34 (PR8) | 120 | genome-wide RNAi | endosomal acidification, vesicular trafficking, mitochondrial metabolism and RNA splicing |
| Konig et al, 2010 | A549 | A/WSN/33 with HA sequence substituted by Renilla. | 295 | genome-wide RNAi | kinase regulated signaling, ubiquitination and phosphatase activity |
| Karlas et al, 2010 | A549 | A/WSN/33 | 287 | genome-wide RNAi | spliceosome, ribosome, nuclear pore complex, ATPase complex, COPI and EIF3. |
| Hao et al, 2008 | D-Mel2 | A/WSN/33 w with HA substituted by VSV-G and NA by Renilla. | 104 | genome-wide RNAi | ATP6V0D1, COX6A1 and NXF1 |
| Shapira et al, 2009 | HBEC | A/PR/8/34 (PR8) | 616 | genome-wide RNAi | RNA binding proteins, apoptosis, MAPK pathway, WNT and NF-kB signaling |
| Song et al, 2011 | SF9 | A/PR/8/34 (H1N1) containing A/Indonesia/5/2005 (H5N1) HA | 37 | virus-like particles proteome | cytoskeleton, translation, chaperone and metabolism |
| Shaw et al, 2008 | Vero | A/WSN/33 | 36 | viral particle proteome | cytoskeletal proteins, annexins, glycolytic enzymes, and tetraspanins |
| Sui et al, 2009 | MDCK | A/Udorn/72 | 110 | Random Homozygous Gene Perturbation | nucleic acid and protein metabolism and intracellular protein trafficking |
| HPIDB (Kumar et al, 2010) | not specified | A/PR/8/34 (H1N1) | 421 | database retrieval | translation, RNA processing, chromatin organization, glycolysis |

**Results**

**Overview of Influenza infection host dependency factors detected by all genome-wide screens.**

In the nine (3-11) genome-wide screens used in this meta-analysis a total of 1574 human genes (table SI) were identified for their importance in Influenza infection, representing over 4.6% of human protein-coding genes (33,868 protein-coding genes from human genome assembly GRCh37.p10).

Analyzing the overlap among genes called by different screens we observe great disparity. No gene was identified by all screens and most genes were identified only in a single list (1436), as seen on figure 2. The seven highest confidence genes (ARCN1, ATP6AP1, COPG, NXF1, RPS10, RPS16 and TUBB) were identified by 4 screens simultaneously. Followed by 20 genes identified by 3 screens (ATP6V0C, ATP6V0D1, CD81, COPA, COPB2, EIF3G, FAU, GAPDH, HSP90AA1, KPNB1, NUP98, PGD, PTBP1, RPS20, RPS24, RPS27A, RPS3, RPS4X, RPS5, WDR18) and 120 genes present in at least 2 screens (figure 3).

The possibility of false positives in these high-throughput screens should always be considered. However, it's also possible that the differences in experimental designs may account for the inconsistencies obtained. This hypothesis might be substantiated by the pairwise analysis of the gene lists (figure 2). The most similarity is observed between the results obtained by Konig, et al (4) and Karlas, et al (5), these two studies used genome-wide RNAi in A549 cells infected with the same Influenza strain (A/WSN/33) as seen in table I. Another interesting aspect revealed by pairwise analysis of the gene lists is that among the ten most similar pairs of lists six of them refer to HPIDB containing comparison pairs, indicating that most screens recapture several previously known Influenza - host interactions already present in available literature curated databases.

**Analysis of high confidence Influenza dependency host factors.**

Genes identified in 2 or more screens were considered high confidence influenza dependency host factors, resulting in a total of 147 genes (figure 3). A protein-protein interaction (PPI) network containing these 147 genes was constructed based on known PPIs from curated databases (13, 14).

Graph theoretic clustering analysis of the above described network with the MCODE algorithm (15) identified seven highly interconnected protein complexes. Genes present in each of these complexes displayed statistically significant overrepresentation of specific gene ontology (GO) categories, as seen in figure 4.

**Figure 2: Overlap of genes detected in different genome-wide screens.** Comparisons of common genes among all gene lists are summarized in the diagram above. Gene lists been compared are indicated by black squares, the number of genes common to all compared lists is indicated in the right-most column. Values in grey indicate genes present only in a single list, marked by a grey square. Different types of experiments are color coded (pink – siRNA screen; blue – RHGP; orange – database; green – virion proteomics).

| | Brass et al, 2009 | Karlas et al, 2010 | Konig et al, 2010 | Hao et al, 2008 | Shapira et al, 2009 | Sui et al, 2009 | HPIDB (Kumar et al, 2010) | Shaw et al, 2008 | Song et al, 2011 |
|---|---|---|---|---|---|---|---|---|---|
| ARCN1 | ● | ● | ● | ● | | | | | |
| ATP6AP1 | ● | ● | ● | ● | | | | | |
| COPG | ● | ● | ● | ● | | | | | |
| NXF1 | ● | ● | | ● | ● | | | | |
| RPS10 | | ● | ● | ● | | | ● | | |
| RPS16 | ● | ● | | ● | | | ● | | |
| TUBB | | | ● | | | | ● | ● | ● |
| ATP6V0C | | ● | ● | ● | | | | | |
| ATP6V0D1 | | ● | ● | ● | | | | | |
| CD81 | | ● | ● | ● | | | | ● | |
| COPA | ● | ● | ● | ● | | | | | |
| COPB2 | ● | ● | ● | | | | | | |
| EIF3G | ● | ● | ● | ● | | | ● | | |
| FAU | | ● | ● | ● | | | ● | | |
| GAPDH | | | | | | | ● | ● | ● |
| HSP90AA1 | | | ● | | | | ● | | ● |
| KPNB1 | | ● | ● | | | | ● | | |
| NUP98 | ● | ● | | | | | | | |
| PGD | | ● | | ● | ● | | | | |
| PTBP1 | ● | | | | | ● | ● | | |
| RPS20 | | | ● | ● | | | ● | | |
| RPS24 | | ● | | | | | ● | | ● |
| RPS27A | | ● | | | | | | | ● |
| RPS3 | | ● | | ● | | | | | |
| RPS4X | ● | ● | | ● | | | | | |
| RPS5 | | ● | ● | ● | | | | | |
| WDR18 | | ● | | | | | ● | | |
| ACTB | | | | | | | | ● | ● |
| AIG1 | | ● | ● | | | | | | |
| AKT1 | | ● | ● | | | ● | | | |
| ATF1 | ● | | | | | | | | |
| ATP5B | | | | ● | | | ● | | |
| ATP6AP2 | | ● | ● | | | | | | |
| ATP6V0B | | ● | ● | | | | | | |
| ATP6V1A | | ● | ● | | | | | | |
| ATP6V1B2 | | ● | ● | | | | | | |
| BARHL2 | | ● | | ● | | | | | |
| BZRAP1 | | ● | ● | | | | | | |
| C14orf166 | ● | | | | | | ● | | |
| C21orf33 | | | | | ● | ● | | | |
| CALCOCO2 | ● | | | | | | | | |
| CFL1 | | | | | | | ● | ● | |
| CLIC4 | | | | | | ● | | | |
| CLK1 | | ● | ● | | | | | | |
| COPB1 | | ● | ● | | | | | | |
| DCLK2 | | ● | ● | | | | | | |
| DDX17 | | | | | ● | | ● | | |
| EEF1A1 | | ● | | | | | ● | | |
| EIF2AK2 | | ● | ● | | | | | | |
| EIF3C | | ● | | ● | | | | | |
| EIF3D | | | | ● | | | ● | | |
| EIF3F | | ● | | ● | | | | | |
| EIF4A2 | ● | | | ● | | | | | |
| EIF5A | | | | | | | ● | ● | ● |
| ENO1 | | | | | | | ● | ● | |
| FAM135A | | ● | ● | | | | | | |
| FASN | | | | | ● | | | ● | |
| FBXW2 | | ● | | | | | | | |

| | Brass et al, 2009 | Karlas et al, 2010 | Konig et al, 2010 | Hao et al, 2008 | Shapira et al, 2009 | Sui et al, 2009 | HPIDB (Kumar et al, 2010) | Shaw et al, 2008 | Song et al, 2011 |
|---|---|---|---|---|---|---|---|---|---|
| FGFR2 | | | ● | | | | | | |
| FKBP5 | | | | | ● | ● | | | |
| FKBP8 | | ● | ● | | | | | | |
| FUS | ● | | | | | | ● | | |
| GRK6 | | | ● | | | | | ● | |
| HAND2 | | | ● | ● | | | | | |
| HIST1H1T | | | | | | | ● | | |
| HIST1H2BN | | ● | | | | | | | |
| HLA-A | | | | | ● | | ● | | |
| HNRNPU | ● | | | | | | ● | | |
| HSP90AB1 | | | | | | | ● | | ● |
| HSPA8 | | | | ● | | | ● | | |
| HSPD1 | | ● | | | | | ● | | |
| IFNGR2 | | | | | ● | | ● | | |
| IGSF1 | | ● | ● | | | | | | |
| IL17RA | | ● | ● | | | | | | |
| ISG15 | | | | | | | | | ● |
| JUN | | ● | ● | | | | | | |
| KRTCAP2 | | | | | ● | | | | |
| LGALS3BP | | | | | ● | | ● | | |
| LRP1B | | ● | ● | | | | | | |
| LY6G6C | | ● | ● | | | | | | |
| MAGEA11 | | | | | ● | | | | |
| MAP2K3 | | ● | ● | | | | | | |
| MDM2 | | ● | ● | | | | | | |
| MRPL12 | | | | ● | | | ● | | |
| MYC | | | | | ● | | | | |
| NHP2L1 | ● | | | | | | | | |
| NTSR1 | | ● | ● | | | | | | |
| NUP153 | | ● | ● | | | | | | |
| NUP205 | ● | ● | | | | | | | |
| OSMR | | ● | ● | | | | | | |
| PAICS | | ● | | | | | ● | | |
| PCID2 | ● | | | ● | | | | | |
| PEPD | ● | | | ● | | | | | |
| PGK1 | | | | | | | ● | ● | |
| PHF2 | | ● | ● | | | | | | |
| PIK3CB | ● | | ● | | | | | | |
| PIK3R2 | | | | | ● | | ● | | |
| PIK3R3 | | | | | ● | | | | |
| PIN1 | | | | | | | | | ● |
| PLAU | | ● | ● | | | ● | | | |
| PLK3 | | ● | ● | | | | | | |
| PPP1R14D | | ● | ● | | | | | | |
| PRPF8 | ● | ● | | | | | | | |
| PSENEN | | ● | ● | | | | | | |
| PTMA | | ● | ● | | | | ● | | |
| PTPRN | | ● | ● | | | | | | |
| PTS | ● | | | | | | | | |
| RAB10 | ● | | | ● | | | | | |
| RAB5A | ● | | | ● | | | | | |
| RACGAP1 | | ● | ● | | | | | | |
| RBCK1 | | | | ● | | | | | |
| RPA1 | | | | | | | ● | ● | |
| RPL15 | | | | ● | | | ● | | |
| RPL23A | | | | ● | | | ● | | |
| RPL28 | | | | ● | | | ● | | |
| RPL35 | ● | | | | | | | | |

| | Brass et al, 2009 | Karlas et al, 2010 | Konig et al, 2010 | Hao et al, 2008 | Shapira et al, 2009 | Sui et al, 2009 | HPIDB (Kumar et al, 2010) | Shaw et al, 2008 | Song et al, 2011 |
|---|---|---|---|---|---|---|---|---|---|
| RPLP1 | | | | | | | ● | | ● |
| RPLP2 | | | | ● | | | | | ● |
| RPS11 | | | | ● | | | ● | | |
| RPS14 | | ● | | ● | | | | | |
| RPS18 | | | | ● | | | ● | | |
| RPS2 | | | | ● | | | ● | | |
| RUNX1 | ● | ● | | | | | | | |
| SF3A1 | | ● | ● | | | | | | |
| SF3B1 | | ● | ● | | | | | | |
| SLC1A3 | ● | | | | ● | | | | |
| SNRPD3 | | | | | | | ● | | |
| SNRPF | | ● | | | ● | | ● | | |
| SNW1 | | ● | | | | ● | | | |
| STARD5 | ● | | | | ● | | | | |
| STX5 | | | | ● | | | ● | | |
| SUPT6H | | ● | | | ● | | | | |
| THOC4 | | | | ● | | | ● | | |
| TNK2 | | ● | ● | | | | | | |
| TRIM25 | | | | | ● | | ● | | |
| TRIM27 | | | | | ● | | ● | | |
| TRIM28 | ● | | ● | | | | | | |
| TUBA1A | | | | | | | ● | ● | |
| TUBA1B | | | | | | | ● | | ● |
| TUBB2C | | | | | | | ● | | ● |
| TUBB3 | | | ● | | | | | | ● |
| UBA52 | | | | | | | ● | | ● |
| UBB | | | | | | | ● | ● | |
| UBC | | | | | ● | | ● | | |
| XPO1 | | ● | | | | | ● | | |
| YWHAZ | | | | | | | ● | | ● |
| ZMAT4 | | | | | ● | | ● | | |

**Figure 3:** Influenza host dependency factors detected in multiple genome-wide screens. Presence – absence map of genes detected in at least two different screens used in this meta-analysis (black – absent, green – present).

This indicates that the identified clusters are likely protein complexes that participate in the enriched biological processes identified. In this case: translation, translation initiation, RNA processing, vATPase complex, Golgi vesicle coating, nuclear pore complex transport, nitrogen compound metabolism and phosphoinositide 3-kinase complex. This analysis identified with high confidence the biological complexes in the host cell that were important for Influenza infection (figure 4).

**Expanding analysis to all identified Influenza dependency host factors.**

In order to expand our analysis to all genes identified in all different screens a new PPI network was created. Nodes in this network represented all genes present in all gene lists used in this meta-analysis. Interaction information was obtained from curated databases (13, 14) in conjunction with the interactions described at HPIDB (11).
Clustering and GO annotation enrichment analysis of this network identified the same clusters found in the high confidence network from figure 4, however expanding the identi-fied complexes to incorporate novel nodes identified in only one screen (figures 5 and 6). Several new complexes with different biological functions were also identified in the analysis of the complete network, bringing the total of biological complexes important for Influenza infection to 17, as seen in figure 5.

**Host cell protein complexes and biological processes involved in the Influenza virus life cycle.**

The host-pathogen interaction network displayed in figure 5 was constructed using all genes present in this meta-analysis (table SI) and curated databases of PPIs (11, 13, 14). The resulting network contained 1,295 nodes connected by 18,452 edges. Clustering analysis identified 17 protein complexes in this network, comprised of 369 nodes highly interconnected by 9,893 edges (926 nodes remained unclustered). These protein complexes are enriched for specific GO categories, indicating their function in the cell (figure 6).
Insights into the role these protein complexes might play during infection can be obtained by correlating them to the Influenza virus life cycle. This life cycle can be summarized into the following: viral entry into the host cell, import of vRNPs into the host nucleus, transcription and replication of the viral genome, export of vRNPs from the host nucleus, and assembly and budding at the host cell plasma membrane.

**Viral entry into the host cell and uncoating**
Viral entry into the cell is mediated by the interaction of the viral membrane protein HA with sialic acid on the host cell's surface. This binding leads to receptor mediated endocytosis and the viral entry into the host cell in an endosome.
Several of the protein complexes identified as host dependency factors for Influenza infection likely play a role in this stage of the viral life cycle. For example, a transient PI3K activation (fig. 6, cluster 5) due to virion attachment promotes its internalization (16), while at the same time inhibiting apoptosis in the early stages of viral infection (17). Regulation of apoptosis (detected in cluster 10, fig. 6) is a common theme during the Influenza virus life cycle, with viral proteins promoting its inhibition in the initial stages of infection and its induction in later stages(18).
Endocytosis of Influenza virus also utilizes pathways used by growth factor receptors (fig. 6, cluster 12), and viral particles are sorted into the same endosomal populations as GFRs. Activation of signaling by these receptors also enhances viral uptake by the cell and later viral replication (19). Protein ubiquitination (fig. 6, cluster 2) is also required for proper sorting of GFRs into the vacuolar pathway, and it is possible this pathway also has an effect in proper sorting of Influenza virus containing vacuoles (19). Inhibition of the ubiquitin-proteasome system has been shown to impair viral entry into the cell by sequestering viruses into endocytic compartments (20).
Once the virus enters the cell through endocytosis, most of its initial trafficking within the host cell is mediated by vesicle transport. The coatomer 1 vesicle transport complex was identified as a host dependency factor in several screens (fig. 6, cluster 9) and it mediates vesicle transport from the early to the late endosome(17). Other vesicle transport factors have also been identified in cluster 15 (fig.6).
The endocytic compartment is also where viral uncoating happens, the endosomal membrane fuses with the viral membrane while the vRNPs are released into the cytoplasm of the host cell. This membrane fusion event requires a low pH environment; this is achieved in the endosome by proton transporting V-type ATPases (clusters 10 and 17, fig. 6). The acidity found in the endosome, created by these vATPases, is also responsible for opening the viral M2 ion transporting channel, thus acidifying the viral core and releasing the vRNPs from M1 so they are free to enter the cytoplasm (21). As described above, 7 out of 17 protein complexes identified as host dependency factors in this meta-analysis are import-ant for these initial steps of Influenza infection. It's possible that this is partially due to the design of the screens used, with 3 of the genome-wide RNAi screens only evaluating the viral life cycle until viral protein translation, therefore focusing only in these initial stages of infection.

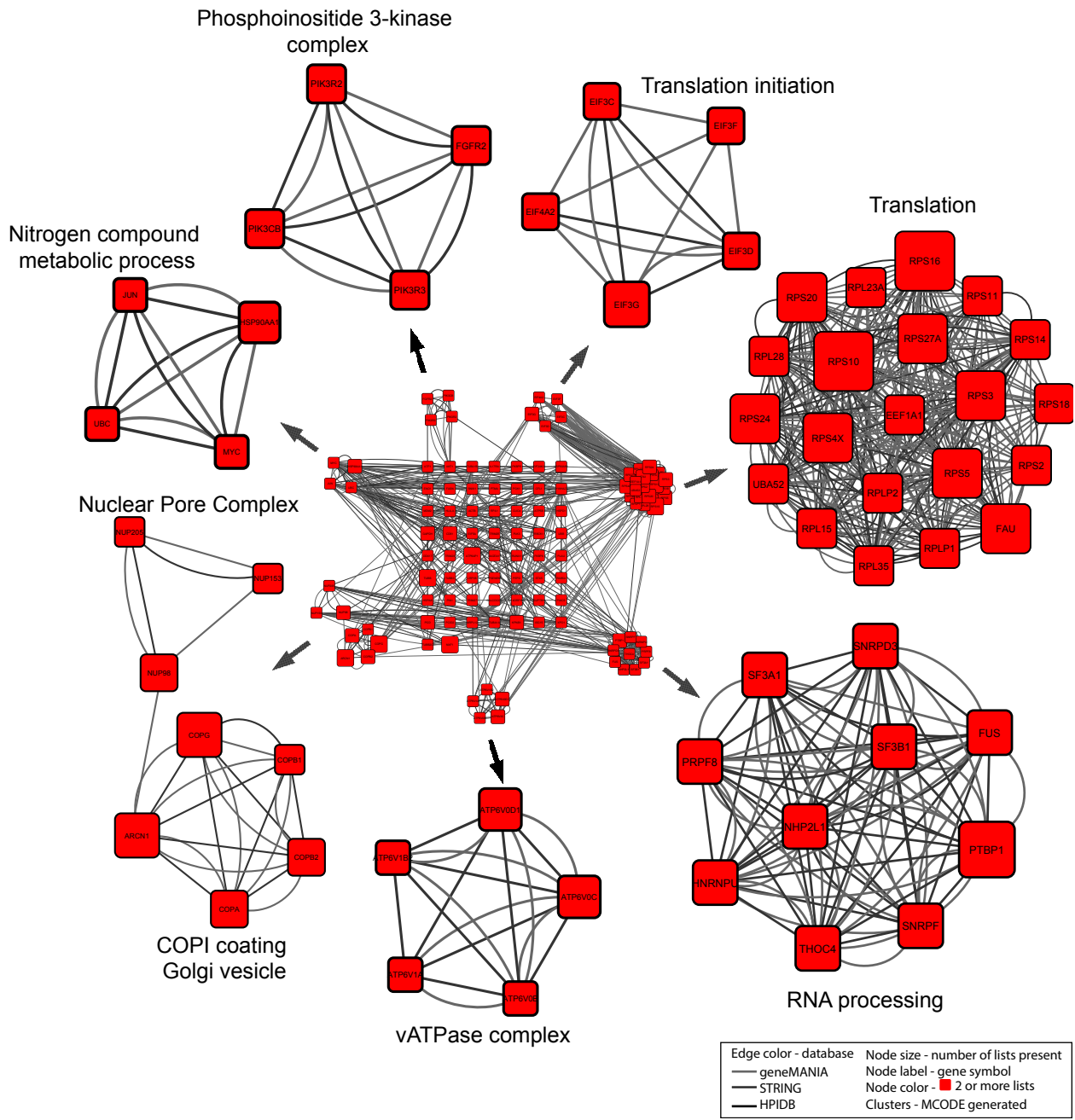

**Figure 4: Highest confidence Influenza dependency host factors interaction network.** Genes identified in two or more screens were used to build the interaction network presented in the center of the figure. MCODE analysis (degree cutoff = 2) identified clusters present, these are shown larger next to the respective network cluster. The main statistically significant GO annotation enrichment in each of the clusters is also indicated.

**Import of vRNPs into the host nucleus**
The vRNPs that have been released into the cytoplasm must enter the nucleus for transcription and replication of the viral genome. Nuclear replication is very beneficial to the virus, as it gains access to the host splicing machinery, facilitates cap-snatching and increases opportunities for evasion of antiviral host responses (22).

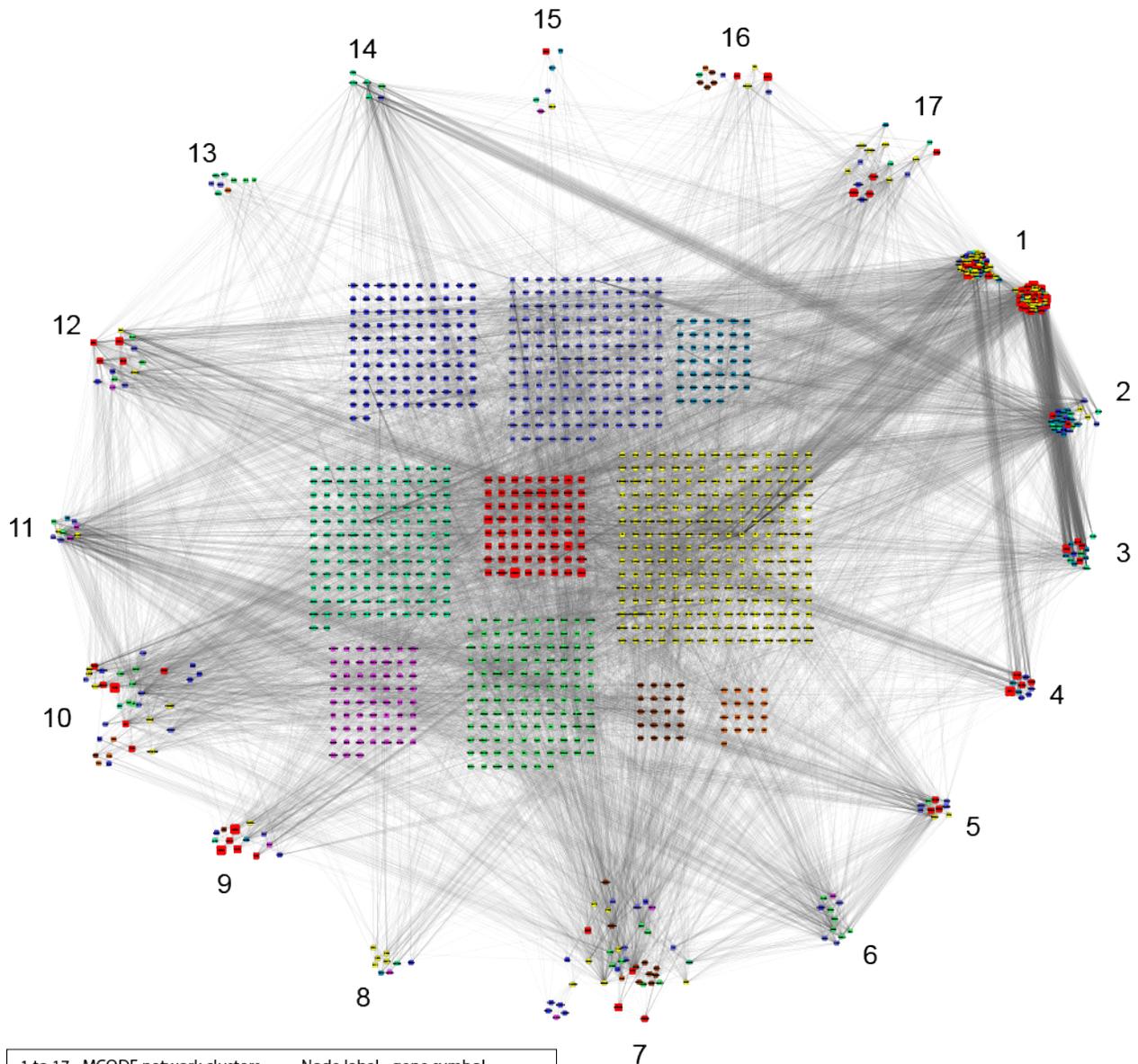

**Figure 5: Influenza dependency host factors interaction network.** Genes identified in all screens were used to build the interaction network displayed. MCODE analysis (degree cutoff = 2) identified 17 clusters present, these are described in more detail on figure 6.

Large structures, such as the vRNPs can only enter the cellular nucleus through translocation across nuclear pore complexes. NPC transport is mediated by the interaction of the target proteins with transport factors known as karyopherins, entrance into the nucleus is mediated by importins while nuclear export is done by exportins. The interaction between target proteins and karyopherins is mediated by short amino acid motifs present in the target proteins; nuclear localization signals (NLSs) mediate interaction with importins while nuclear export signals promote interactions with exportins. Viral proteins present in the vRNP (NP, PA, PB1 and PB2) contain nuclear localization signals (NLS) (21), these NLSs become apparent after the vRNP is released from the M1 viral protein, enabling them to enter the nucleus through nuclear pore complex transport pathways (cluster 4, fig 6). Newly synthesized viral proteins that are part of the vRNPs (PA, PB1, PB2 and NP) also have to be imported into the nucleus after cytoplasmic translation to assemble novel vRNP particles. These proteins interact with a range of importins for nuclear entrance and it has been shown that these importins function as chaperones for these subunits as well as chaperoning the formation of the vRNP complex itself (22).

RNA processing

Translation

Negative regulation of ubiquitination

Translation initiation

Nuclear Pore Complex
RNA transport

Phosphoinositide 3-kinase complex
Signal transmission

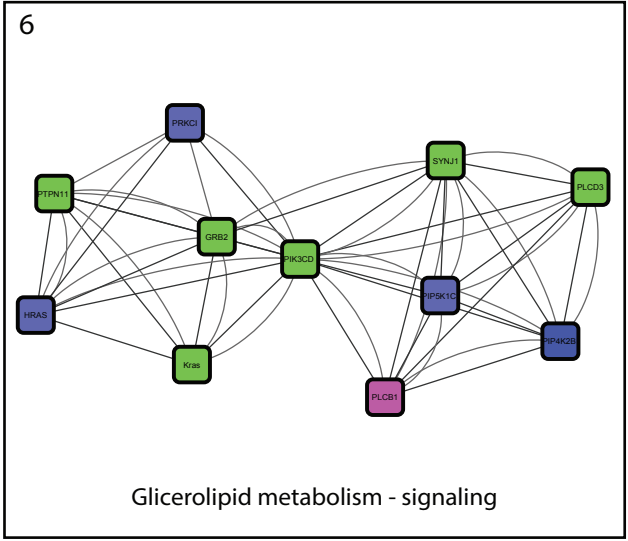

Glicerolipid metabolism - signaling

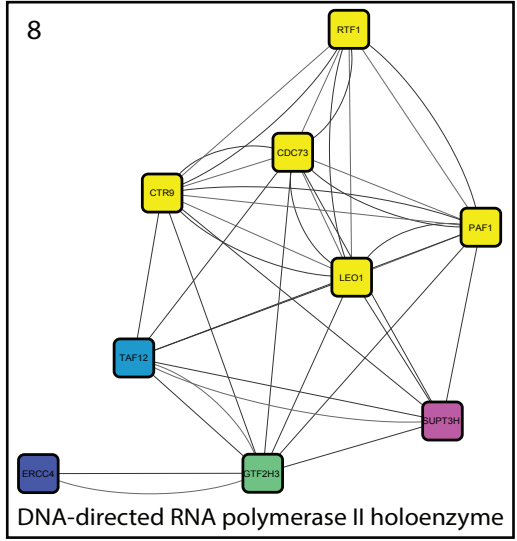

DNA-directed RNA polymerase II holoenzyme

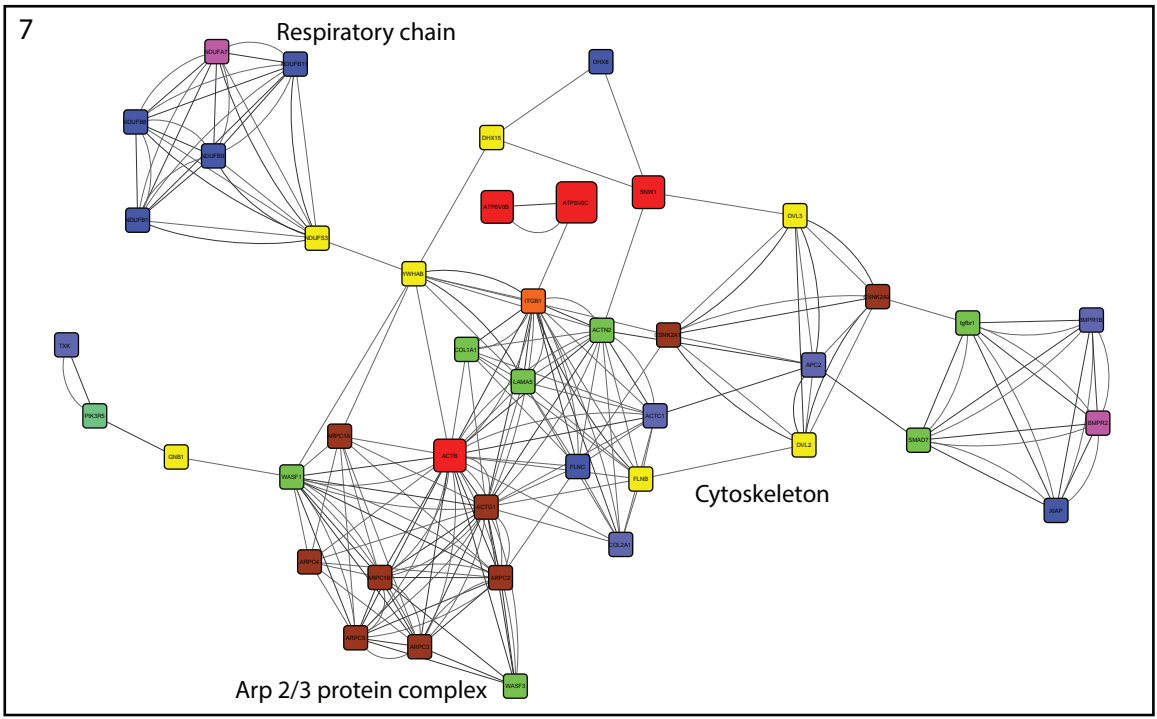

Respiratory chain

Cytoskeleton

Arp 2/3 protein complex

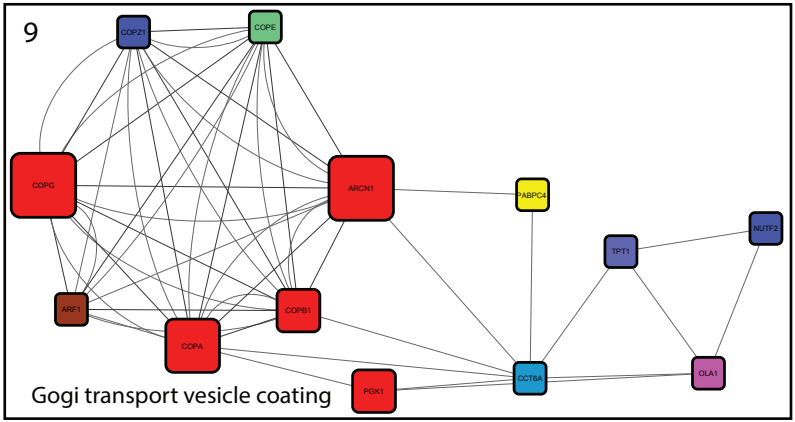

Gogi transport vesicle coating
November 2012     10...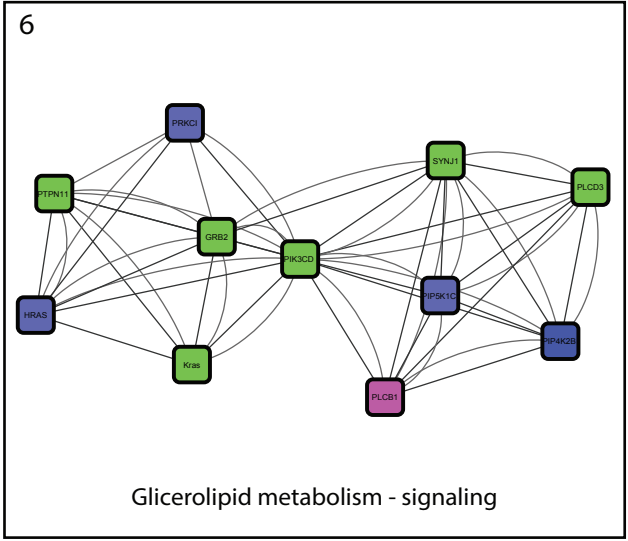

Glicerolipid metabolism - signaling

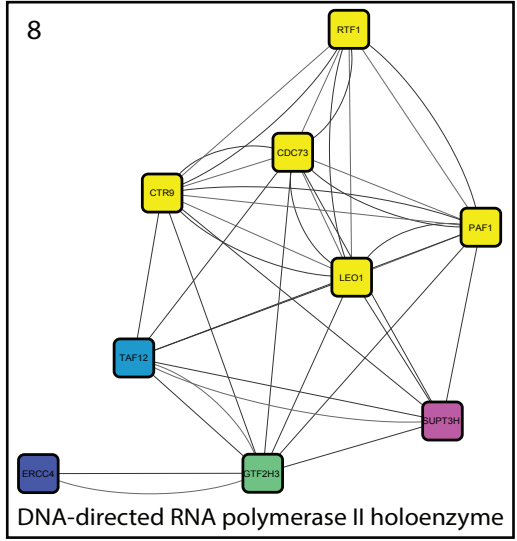

DNA-directed RNA polymerase II holoenzyme

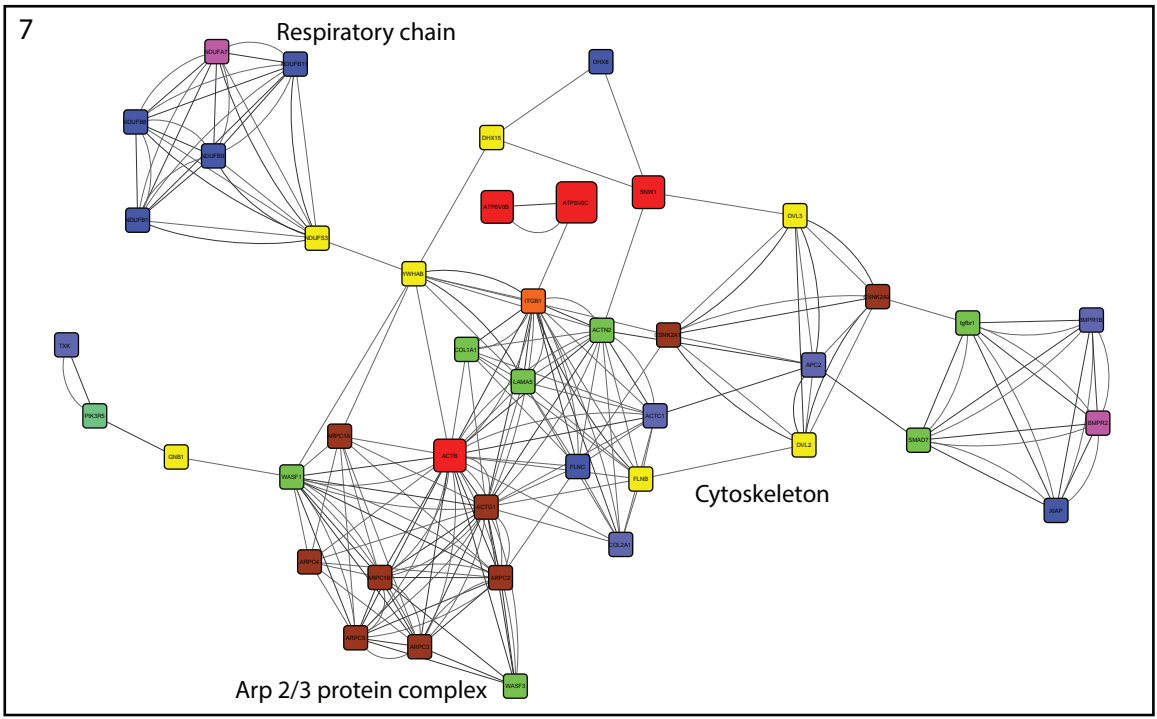

Respiratory chain

Cytoskeleton

Arp 2/3 protein complex

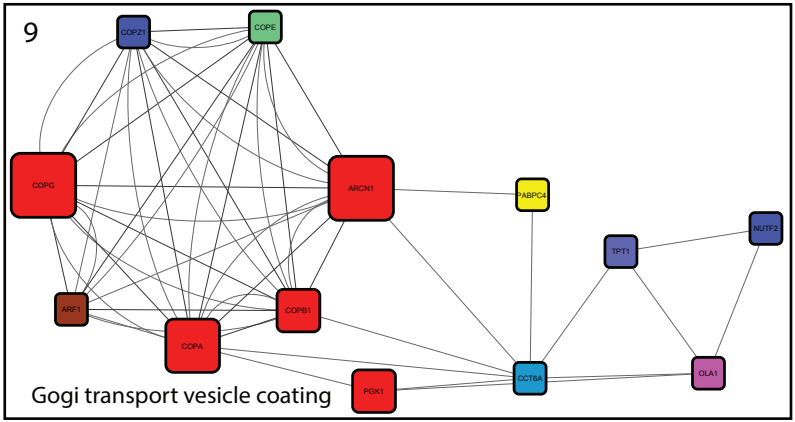

Gogi transport vesicle coating

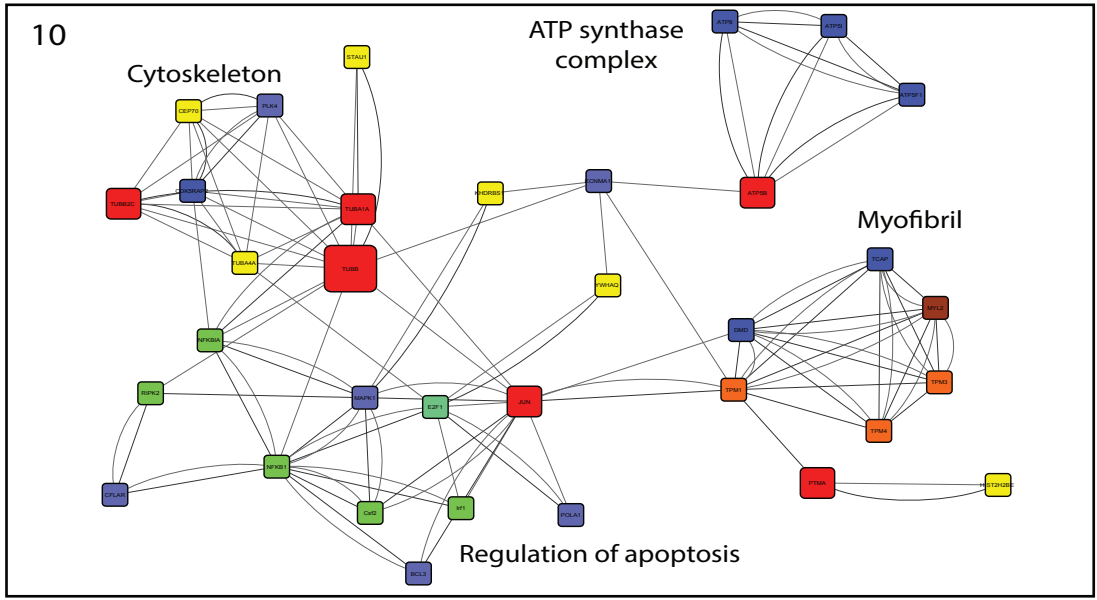
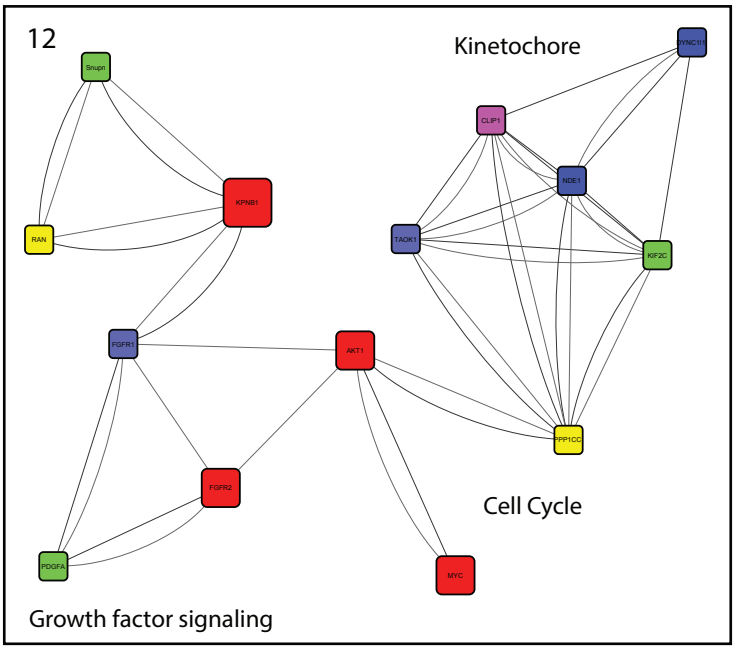
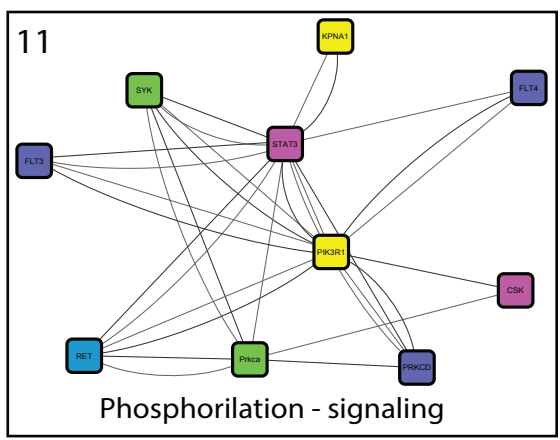
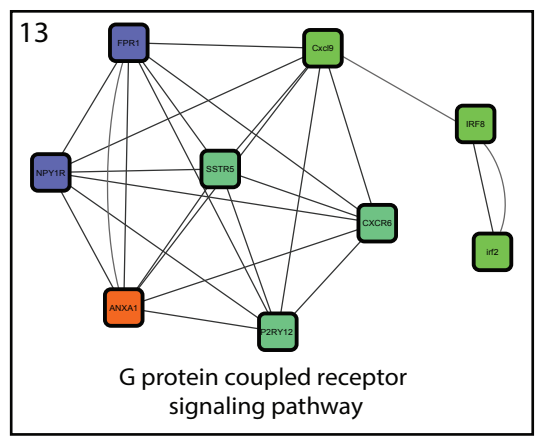

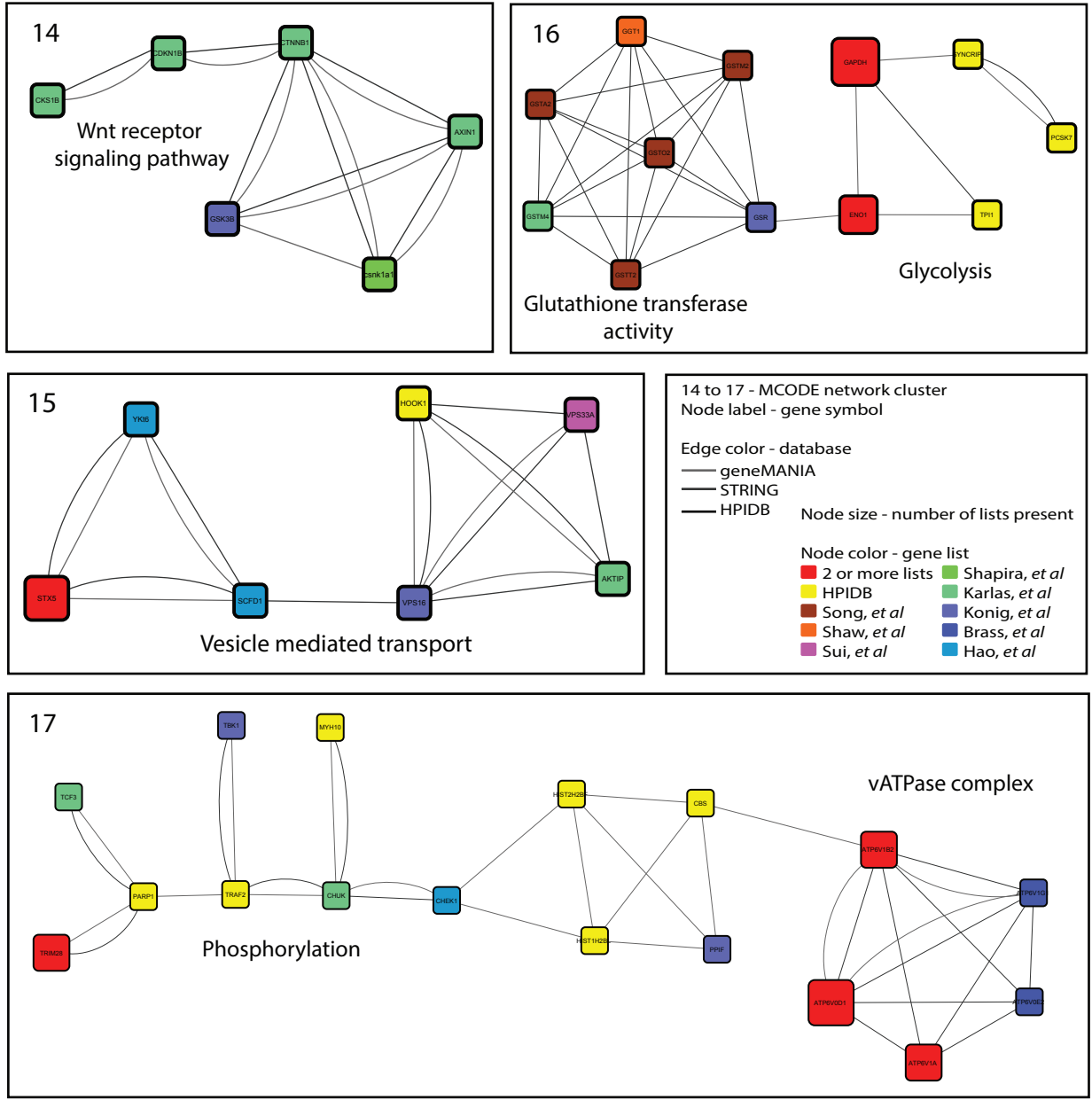

**Figure 6: Gene clusters identified in the PPI network of Influenza host dependency factors.** Gene clusters identified in figure 5 are shown here in more detail. The main statistically significant GO annotation enrichment in each cluster is also indicated.

**Transcription and replication of the viral genome**
Once inside the nucleus the vRNPs get transcribed into mRNAs and replicated via positive sense RNAs. While the viral polymerase complex catalyzes transcription and genome replication, several host proteins play a role in these processes and can affect their efficiency.
Host RNA polymerase II (RNPII) for example (identified in fig. 6, cluster 8), when in its active state (phosphorylated Ser5 on the CTD) can bind the viral polymerase complex (21). RNAPII has also been shown to be required for transcription of viral mRNAs (17).

As mentioned previously, nuclear transcription of viral mRNAs gives it access to the host splicing machinery. The Influenza virus genome has two segments that encode two different proteins due to alternative splicing, segment 7 encodes M1 and M2 while segment 8 encodes NS1 and NS2 (or NEP). The Influenza virus uses the host cell splicing machinery to express these two proteins (cluster 1, fig. 6). At the same time viral proteins prevent the host cell from using its own splicing machinery hampering the processing of host mRNAs (21).

**Nuclear export of vRNPs and viral mRNAs and translation of viral proteins.**

The viral mRNAs encoding viral proteins must be exported from the nucleus for translation. Viral genomes, packaged into vRNPs also have to be exported from the nucleus for assembly of novel viral particles in the cytoplasm of the host cell. This nucleo-cytoplasmic trafficking occurs through transport across nuclear pore complexes, using different transport factors (cluster 4, fig. 6).

The export of viral mRNAs is dependent on NFX1, a known host mRNA export pathway that gets exploited by the Influenza virus. Not only does the virus highjacks this pathway for export of its own mRNAs it also blocks its usage for export of host mRNAs, which are then retained in the nucleus (23). The nuclear retention of host mRNAs increases viral "cap-snatching" and inhibits the host immune response (17). Another important modulator of immune response is WNT signaling (cluster 14, fig. 6), this pathway not only affects the host immune response but also significantly impacts Influenza replication of the viral genome (7). The nuclear export of negative sense vRNPs containing the viral genome is mediated by CRM1. In order for vRNP export to occur M1 and NS2 are required. M1 can interact with the vRNA and with NS2, while NS2 then interacts with CRM1 for nuclear export (21). Interestingly, PI3K signaling (cluster 5, fig. 6) has also been shown to affect vRNP nuclear export, as well as viral mRNA transcription and translation (16).

When the viral mRNAs reach the cytoplasm they require the cellular machinery for translation (cluster 1, fig. 6). Viral infection shuts-off translation of host-proteins and enhances translation of viral proteins. This meta- analysis identified not only several genes involved in mRNA translation (cluster 1, fig. 6), but also a large cluster of translation initiation factors (cluster 3, fig. 6). It's possible that these translation initiation factors play an important role in the selectivity of virus specific translation initiation or on the inhibition of host translation initiation.

**Assembly and budding of viral particles**

During the last stages of the Influenza virus life cycle all viral components are transported to the host cell plasma membrane for assembly. Transport of viral proteins to these sites is dependent on several host factors. Several viral proteins get to the plasma membrane through vesicle transport pathways (cluster 15, fig. 6). The COPI complex has an important role in the transport of viral glicoproteins (cluster 9, fig. 6). The transport of vRNPs to the budding site may be mediated by its interaction with host cytoskeletal components (clusters 7 and 10, fig. 6), facilitated by the interaction of NP and M1 with actin. The last step in Influenza infection, viral budding from the host cell membrane is also dependent on the host cell cytoskeleton (17).

**Other identified functions**

Several other new protein complexes have been identified in this meta-analysis, whose function in the Influenza life cycle is still unknown, such as: cluster 6, glicerolipid metabolism; cluster 7, Arp 2/3 protein complex and respiratory chain; cluster 11, phophorylation signaling; cluster 13, G protein coupled receptor signaling; cluster 16, Glutathione transferase and glycolysis. Further studies are required to determine the relevance of these results and to better define the importance of these host factors in Influenza infection.

## Concluding remarks

As seen in several viruses, Influenza A takes advantage of several components of the host cell machinery for its replication. Recent advances in science have now allowed the development of genome-wide screens that can be used to investigate the importance of nearly all host genes in the viral life cycle.

Taking advantage of such technique, several screens in the past few years have evaluated genome-wide host factors dependencies for Influenza infection. This meta-analysis evaluates nine such screens to identify commonalities that point towards protein complexes that can be identified with high confidence as host dependency factors for Influenza infection.

These high confidence host dependency factors identified could be important drug targets to treat Influenza infection, aiming to alleviate the large burden brought upon global health by this virus.

## Methods

This meta-analysis used results for published genome-wide screens for Influenza virus host dependency factors. The list of dependency factors collected from each screen was based on the reporting paper's final results, after results had been filtered into high confidence gene lists. Gene lists from screens performed in other species were matched to their annotated human orthologs. Screens used are detailed in table I and complete gene list can be found in table SI.

Comparison of gene list was performed using the statistical language R (24) in RStudio (25). Network analysis were done in Cytoscape (26), using the plugins geneMANIA (14), MCODE (15) and BiNGO (27). Interaction network data imported into Cytoscape was also imported from the STRING database (13).

Table SI: Gene lists used in this meta-analysis. Complete list of host factors necessary for Influenza infection identified in each different genome-wide screens.

| Brass et al | | Shapira et al | | Karlas et al | | Konig et al | | Hao et al | | Sui et al | | HPIDB | | Shaw et al | | Song et al | |
|---|---|---|---|---|---|---|---|---|---|---|---|---|---|---|---|---|---|
| Symbol | ID | Symbol | ID | Symbol | ID | Symbol | ID | Symbol | ID | Symbol | ID | Symbol | ID | Symbol | ID | Symbol | ID |
| AHCY | 191 | ACP2 | 53 | ACACA | 31 | ACTC1 | 70 | ARCN1 | 372 | ACAP2 | 23527 | SEPT1 | 1731 | ACTB | 60 | ACTB | 60 |
| ABCD1 | 215 | ACTN2 | 88 | ABCC6 | 368 | ACVR2A | 92 | ATP5B | 506 | ADAM20 | 8748 | ABLIM1 | 3983 | AKR1B1 | 231 | ACTG1 | 71 |
| XIAP | 331 | Adam8 | 101 | ARCN1 | 372 | ADRA1B | 147 | ATP5C1 | 509 | AKT1 | 207 | ACOT9 | 23597 | ANXA1 | 301 | ARF1 | 375 |
| ARCN1 | 372 | ADCY7 | 113 | ASPA | 443 | ADRBK2 | 157 | ATP6V0C | 527 | ANKS1A | 23294 | ACTA2 | 59 | ANXA11 | 311 | ARF6 | 382 |
| ASAH1 | 427 | AMHR2 | 269 | ATP1A2 | 477 | AKT1 | 207 | ATP6AP1 | 537 | ARID1B | 57492 | ACTG2 | 72 | ANXA2 | 302 | ARPC1A | 10552 |
| ATF1 | 466 | ATF4 | 468 | ATP6V1A | 523 | ANPEP | 290 | CCT6A | 908 | BACH1 | 571 | ADAR | 103 | ANXA4 | 307 | ARPC1B | 10095 |
| ATP1A3 | 478 | Cacna1a | 773 | ATP6V1B2 | 526 | APOA1 | 335 | CHEK1 | 1111 | BCL2 | 596 | ALDOA | 226 | ANXA5 | 308 | ARPC2 | 10109 |
| ATP5F1 | 515 | CALM | 805 | ATP6V0C | 527 | APP | 351 | COX6A1 | 1337 | BMPR2 | 659 | ARIH1 | 25820 | CD59 | 966 | ARPC3 | 10094 |
| ATP5I | 521 | CAMK2G | 818 | ATP6AP1 | 537 | AQP4 | 361 | EIF1AX | 1964 | BPTF | 2186 | ATF1 | 466 | CD81 | 975 | ARPC4 | 10093 |
| ATP6V0B | 533 | CISH | 1154 | B2M | 567 | ARAF | 369 | EIF4A2 | 1974 | BTBD8 | 284697 | ATP5B | 506 | CD9 | 928 | ARPC5 | 10092 |
| ATP6AP1 | 537 | clu | 1191 | OPN1SW | 611 | ARCN1 | 372 | EIF5 | 1983 | C21orf33 | 8209 | ATXN2L | 11273 | CFL1 | 1072 | CSNK2A1 | 1457 |
| BUB1B | 701 | COL1A1 | 1277 | RUNX1 | 861 | ATP6V1A | 523 | FAU | 2197 | C4orf32 | 132720 | BANP | 54971 | CNP | 1267 | CSNK2A2 | 1459 |
| CACNB4 | 785 | copA | 1314 | CD47 | 961 | ATP6V1B2 | 526 | HMGCR | 3156 | CHM | 1121 | BASP1 | 10409 | DBI | 1622 | EIF5A | 1984 |
| RUNX1 | 861 | klf6 | 1316 | CD48 | 962 | ATP6V0C | 527 | HSPA8 | 3312 | CIP29 | 84324 | BCAP29 | 55973 | DSTN | 11034 | FABP3 | 2170 |
| CEACAM7 | 1087 | Atf2 | 1386 | CD58 | 965 | ATP6V0B | 533 | EIF3E | 3646 | CLIP1 | 6249 | BHLHB2 | 8553 | ENO1 | 2023 | GABRR1 | 2569 |
| CLK3 | 1198 | Csf2 | 1437 | CD81 | 975 | ATP6AP1 | 537 | KRAS | 3845 | CLNS1A | 1207 | BLZF1 | 8548 | FASN | 2194 | GAPDH | 2597 |
| CLN5 | 1203 | csnk1a1 | 1452 | CDKN1B | 1027 | BCL3 | 602 | MAN2A2 | 4122 | COX5A | 9377 | C10orf35 | 219738 | GAPDH | 2597 | GLRX | 2745 |
| COPA | 1314 | CYLD | 1540 | CEL | 1056 | BMPR1B | 658 | MAT2A | 4144 | CRLS1 | 54675 | C10orf96 | 374355 | GGT1 | 2678 | GSTA2 | 2939 |
| COPB1 | 1315 | cyp2j2 | 1573 | CHRM1 | 1128 | CAD | 790 | ALDH6A1 | 4329 | CSK | 1445 | C14orf166 | 51637 | GPC4 | 2239 | GSTM2 | 2946 |
| CPN2 | 1370 | Dlx2 | 1746 | CHUK | 1147 | CAMK2B | 816 | CNOT3 | 4849 | CSNKID | no data | C16orf45 | 89927 | HSPB1 | 3315 | GSTO2 | 119391 |
| DHX8 | 1659 | dusp5 | 1847 | CKS1B | 1163 | CAPN6 | 827 | NTSR1 | 4923 | DDX17 | 10521 | C1orf94 | 84970 | ITGB1 | 3688 | GSTT2 | 2953 |
| DMD | 1756 | elk1 | 2002 | AP2M1 | 1173 | CD81 | 975 | NUP98 | 4928 | DDX58 | 23586 | C22orf28 | 51493 | PGK1 | 5230 | HSP90AA1 | 3320 |
| DYNC1I1 | 1780 | EP300 | 2033 | CLK1 | 1195 | CDK4 | 1019 | P4HA1 | 5033 | DEDD2 | 162989 | CALCOCO1 | 57658 | PKM2 | 5315 | HSP90AB1 | 3326 |
| DSC3 | 1825 | EPHA4 | 2043 | CNGB1 | 1258 | CLK1 | 1195 | PEPD | 5184 | DENND1B | 163486 | CALM1 | 801 | PRDX1 | 5052 | HSP90B1 | 7184 |
| EIF2S1 | 1965 | Acsl4 | 2182 | PLK3 | 1263 | PLK3 | 1263 | PGD | 5226 | DPYD | 1806 | CALM2 | 805 | S100A11 | 6282 | MYL2 | 4633 |
| EIF4A2 | 1974 | FASN | 2194 | COPA | 1314 | COL2A1 | 1280 | PLXNA2 | 5362 | DYRK1A | 1859 | CALM3 | 808 | TAGLN | 6876 | PIN1 | 5300 |
| EPRS | 2058 | FGFR2 | 2263 | COPB1 | 1315 | COPA | 1314 | PSMB1 | 5689 | FAM13B1 | 51306 | CALR | 811 | TPM1 | 7168 | PTP4A1 | 7803 |
| ERCC4 | 2072 | FKBP5 | 2289 | CRY2 | 1408 | CREB1 | 1385 | PSMB3 | 5691 | FAM13C1 | 220965 | CBS | 875 | TPM3 | 7170 | PTP4A2 | 8073 |
| FCGR2A | 2212 | GOLGA3 | 2802 | CRYAA | 1409 | CRHR1 | 1394 | PSMB4 | 5692 | FBXL17 | 64839 | CCDC33 | 80125 | TPM4 | 7171 | PTP4A3 | 11156 |
| FOXE1 | 2304 | Gpd2 | 2820 | CTNNB1 | 1499 | CSE1L | 1434 | PSMB6 | 5694 | FHIT | 2272 | CCT5 | 22948 | TUBA1A | 7846 | RPLP1 | 6176 |
| FLNC | 2318 | GRB2 | 2885 | CYP17A1 | 1586 | CSNK1G2 | 1455 | PSMC1 | 5700 | FKBP5 | 2289 | CDC42EP4 | 23580 | TUBB | 203068 | RPLP2 | 6181 |
| FUS | 2521 | HLA-A | 3105 | DBT | 1629 | CTSG | 1511 | PSMC3 | 5702 | GLCE | 26035 | CDC73 | 79577 | UBB | 7314 | RPS24 | 6229 |
| GJA3 | 2700 | HLA-G | 3135 | DLG2 | 1740 | CTSW | 1521 | PSMD11 | 5717 | GNL3L | 54552 | CEP70 | 80321 | UCHL1 | 7345 | RPS27A | 6233 |
| GSK3A | 2931 | NR4A1 | 3164 | E2F1 | 1869 | DAPK3 | 1613 | PSMD12 | 5718 | GOLGA4 | 2803 | CFL1 | 1072 | WDR1 | 9948 | SYBL1 | 6845 |
| HIST1H1T | 3010 | ICAM1 | 3383 | EEF1A1 | 1915 | DHCR7 | 1717 | RAB1A | 5861 | GON4L | 54856 | CHD1 | 1105 | | | TRUB2 | 26995 |
| HNRNPU | 3192 | IRF8 | 3394 | EPHB6 | 2051 | DIO1 | 1733 | RAB5A | 5868 | GRK6 | 2870 | CHMP1B | 57132 | | | TUBA1B | 10376 |
| LFNG | 3955 | Ifit2 | 3433 | FAU | 2197 | TRDMT1 | 1787 | DPF2 | 5977 | HEATR7A | 727957 | CHMP6 | 79643 | | | TUBA1C | 84790 |
| LRPAP1 | 4043 | IFNGR2 | 3460 | GDF10 | 2662 | DSP | 1832 | RET | 5979 | HERC6 | 55008 | CKB | 1152 | | | TUBA3C | 7278 |
| SH2D1A | 4068 | IGFBP6 | 3489 | GCLC | 2729 | DUSP3 | 1845 | RPL5 | 6125 | IQGAP1 | 8826 | CMTM5 | 116173 | | | TUBA3D | 113457 |
| MFAP1 | 4236 | IL15RA | 3601 | GNRH2 | 2797 | MARK2 | 2011 | RPL15 | 6138 | KDSR | 2531 | COL4A3BP | 10087 | | | TUBB | 203068 |
| MPG | 4350 | Irak1 | 3654 | GRIN2C | 2905 | ENG | 2022 | RPL23A | 6147 | KIAA1033 | 23325 | CPSF1 | 29894 | | | TUBB1 | 81027 |
| ATP8 | 4509 | Irf1 | 3659 | GSTM4 | 2948 | EPHA7 | 2045 | RPL28 | 6158 | KIAA1107 | 23285 | CPSF4 | 10898 | | | TUBB2A | 7280 |
| MYBL2 | 4605 | irf2 | 3660 | GTF2H3 | 2967 | EPHB2 | 2048 | RPLP2 | 6181 | KIDINS220 | 57498 | CREB3 | 10488 | | | TUBB2B | 347733 |

| | | | | | | | | | | | | | | | | |
|---|---|---|---|---|---|---|---|---|---|---|---|---|---|---|---|---|
| MYO1E | 4643 | irf6 | 3664 | HPGD | 3248 | EPHB4 | 2050 | MRPL12 | 6182 | LRRC16A | 55604 | CRX | 1406 | | | TUBB2C | 10383 |
| NAGA | 4668 | Itpkb | 3707 | HSF4 | 3299 | F13A1 | 2162 | RPS2 | 6187 | LSM4 | 25804 | CRYAB | 1410 | | | TUBB3 | 10381 |
| NDUFB1 | 4707 | junb | 3726 | HSPD1 | 3329 | FGFR1 | 2260 | RPS3 | 6188 | MARK3 | 4140 | CSDA | 8531 | | | TUBB4 | 10382 |
| NDUFB8 | 4714 | Kras | 3845 | TNC | 3371 | FGFR2 | 2263 | RPS3A | 6189 | MBOAT2 | 129642 | CTR9 | 9646 | | | TUBB8 | 347688 |
| NDUFB9 | 4715 | LAMA5 | 3911 | IFNA7 | 3444 | FGFR4 | 2264 | RPS4X | 6191 | MDN1 | 23195 | DDB1 | 1642 | | | UBA52 | 7311 |
| NHP2L1 | 4809 | LGALS3BP | 3959 | IFNAR2 | 3455 | FLT3 | 2322 | RPS5 | 6193 | MRPL42 | 28977 | DDX17 | 10521 | | | VAMP1 | 6843 |
| NOS3 | 4846 | lgals9 | 3965 | IFNGR2 | 3460 | FLT4 | 2324 | RPS6 | 6194 | N4BP2 | 55728 | DDX3X | 1654 | | | VAMP2 | 6844 |
| NUP88 | 4927 | lmna | 4000 | IGSF1 | 3547 | AFF2 | 2334 | RPS8 | 6202 | NDRG3 | 57446 | DDX3Y | 8653 | | | VAMP3 | 9341 |
| NUP98 | 4928 | Smad5 | 4090 | IK | 3550 | FNTB | 2342 | RPS9 | 6203 | NDUFA7 | 4701 | DHX15 | 1665 | | | VAMP4 | 8674 |
| PCSK5 | 5125 | SMAD7 | 4092 | IL1A | 3552 | FOLH1 | 2346 | RPS10 | 6204 | NDUFAF2 | 91942 | DHX30 | 22907 | | | VAMP5 | 10791 |
| PGD | 5226 | MAGEA11 | 4110 | IRF2 | 3660 | FPR1 | 2357 | RPS11 | 6205 | NOTUM | 147111 | DHX9 | 1660 | | | VAMP8 | 8673 |
| PIGH | 5283 | Mfap1 | 4236 | JUN | 3725 | FRK | 2444 | RPS13 | 6207 | NPNT | 255743 | DKFZp686D0972 | 345651 | | | VPS28 | 51160 |
| PIP | 5304 | Cxcl9 | 4283 | KCNJ12 | 3768 | FRAP1 | 2475 | RPS14 | 6208 | OLA1 | 29789 | DOCK8 | 81704 | | | YWHAZ | 7534 |
| PLP1 | 5354 | MYC | 4609 | KNG1 | 3827 | G6PD | 2539 | RPS15A | 6210 | OOG5 | no data | DVL2 | 1856 | | | ZNF347 | 84671 |
| PPARA | 5465 | NBN | 4683 | KIF11 | 3832 | GABBR1 | 2550 | RPS16 | 6217 | PAPOLA | 10914 | DVL3 | 1857 | | | | |
| PRCC | 5546 | Neu1 | 4758 | KPNB1 | 3837 | GAK | 2580 | RPS17 | 6218 | PC | 5091 | DYNLL2 | 140735 | | | | |
| PRSS8 | 5652 | NF1 | 4763 | LGALS2 | 3957 | GJA8 | 2703 | RPS18 | 6222 | PCMT1 | 5110 | EEF1A1 | 1915 | | | | |
| PTBP1 | 5725 | NFE2L1 | 4779 | MAN2B1 | 4125 | GRK5 | 2869 | RPS20 | 6224 | PEX5L | 51555 | EEF1A2 | 1917 | | | | |
| PTBS1 | 5725 | NFIL3 | 4783 | MATN3 | 4148 | GRK6 | 2870 | RRBP1 | 6238 | PIK3C3 | 5289 | EEF2 | 1938 | | | | |
| PTPRA | 5786 | NFKB1 | 4790 | MDM2 | 4193 | GSK3B | 2932 | SCD | 6319 | PKD2L2 | 27039 | EIF2AK2 | 5610 | | | | |
| PTS | 5805 | NFKBIA | 4792 | MSRA | 4482 | GSR | 2936 | TAF12 | 6883 | PLAU | 5328 | EIF2B4 | 8890 | | | | |
| PVALB | 5816 | Nfkbie | 4794 | MST1R | 4486 | HRAS | 3265 | TDG | 6996 | PLCB1 | 23236 | EIF2C1 | 26523 | | | | |
| RAB5A | 5868 | OAS1 | 4938 | MYC | 4609 | HSP90AA1 | 3320 | UQCRC2 | 7385 | PLEC1 | 5339 | EIF2S2 | 8894 | | | | |
| RNH1 | 6050 | pck2 | 5106 | MYOD1 | 4654 | HTR2A | 3356 | VCP | 7415 | POLK | 51426 | EIF3D | 8664 | | | | |
| RPS4X | 6191 | PDGFA | 5154 | NAIP | 4671 | IGSF1 | 3547 | ZNF84 | 7637 | PRKD3 | 23683 | EIF3G | 8666 | | | | |
| RPS16 | 6217 | PGD | 5226 | NPAS1 | 4861 | IL5RA | 3568 | EIF3B | 8662 | PTBP1 | 5725 | EIF5A | 1984 | | | | |
| ST8SIA1 | 6489 | pik3c2a | 5286 | NTHL1 | 4913 | IL9R | 3581 | EIF3C | 8663 | RAD51L1 | 5890 | EIF5A2 | 56648 | | | | |
| SLC1A3 | 6507 | PIK3CD | 5293 | NUP98 | 4928 | ITGA2B | 3674 | EIF3D | 8664 | REV3L | 5980 | EIF5AL1 | 143244 | | | | |
| SLC2A2 | 6514 | PIK3R2 | 5296 | SERPINB2 | 5055 | ITGA3 | 3675 | EIF3F | 8665 | RFNG | 5986 | ELAVL1 | 1994 | | | | |
| SNRPB | 6628 | Prkca | 5578 | PDGFRL | 5157 | JAK2 | 3717 | EIF3G | 8666 | RFX3 | 5991 | ENO1 | 2023 | | | | |
| SNRPD2 | 6633 | Mapk11 | 5600 | PEPD | 5184 | JUN | 3725 | EIF3H | 8667 | RGNEF | 64283 | ENY2 | 56943 | | | | |
| SNRPD3 | 6634 | MAPK13 | 5603 | PHF2 | 5253 | KCNJ3 | 3760 | RNMT | 8731 | RGS2 | 5997 | ERH | 2079 | | | | |
| SRP54 | 6729 | Map2k3 | 5606 | SERPINA1 | 5265 | KCNJ11 | 3767 | ATP6V0D1 | 9114 | RPA1 | 6117 | EXOSC8 | 11340 | | | | |
| THRSP | 7069 | PRKY | 5616 | PIK3C2G | 5288 | KCNMA1 | 3778 | HAND2 | 9464 | SEMA3A | 10371 | EZR | 7430 | | | | |
| TPSAB1 | 7177 | psmb9 | 5698 | PIK3CB | 5291 | KPNB1 | 3837 | PSMD6 | 9861 | SF3A3 | 10946 | FAU | 2197 | | | | |
| TSTA3 | 7264 | PTPN11 | 5781 | PIN1 | 5300 | LIMK1 | 3984 | AMMECR1 | 9949 | SLC25A25 | 114789 | FBL | 2091 | | | | |
| TUFT1 | 7286 | RBBP6 | 5930 | PLAU | 5328 | LTK | 4058 | NUP153 | 9972 | SLC30A9 | 10463 | FLNB | 2317 | | | | |
| UBE2A | 7319 | TRIM27 | 5987 | PLD2 | 5338 | MDM2 | 4193 | THOC4 | 10189 | SLIRP | 81892 | FUS | 2521 | | | | |
| UBE2E1 | 7324 | Rnase4 | 6038 | POLR2H | 5437 | MAP3K11 | 4296 | TIMM17B | 10245 | SNW1 | 22938 | FXR2 | 9513 | | | | |
| ZNF16 | 7564 | RNASEL | 6041 | POLR2L | 5441 | NHP2L1 | 4809 | STUB1 | 10273 | SOSTDC1 | 25928 | GABPB2 | 2553 | | | | |
| ZNF132 | 7691 | RXRG | 6258 | MAP2K3 | 5606 | NPY1R | 4886 | NXF1 | 10482 | SPINT2 | 10653 | GAPDH | 2597 | | | | |
| ZNF154 | 7710 | CCL3 | 6348 | PRPS1 | 5631 | NTRK1 | 4914 | YKt6 | 10652 | SSH3 | 54961 | GLYAT | 10249 | | | | |
| ZNF224 | 7767 | Sdc4 | 6385 | PSAP | 5660 | NTRK2 | 4915 | RAB10 | 10890 | STAT3 | 6774 | GMCL1 | 64395 | | | | |
| LZTR1 | 8216 | SHC1 | 6464 | PSMA1 | 5682 | ROR2 | 4920 | RNPS1 | 10921 | STIP1 | 10963 | GNB1 | 2782 | | | | |
| TMEM187 | 8269 | SLC1A3 | 6507 | PSMC6 | 5706 | NTSR1 | 4923 | COPG | 22820 | STXBP1 | 6812 | GNB2 | 2783 | | | | |
| FZD8 | 8325 | SLC7A1 | 6541 | PSMD2 | 5708 | PAK2 | 5062 | CLUAP1 | 23059 | SUPT3H | 8464 | GNB3 | 2784 | | | | |

| | | | | | | | | | | | | |
|---|---|---|---|---|---|---|---|---|---|---|---|---|
| PIP4K2B | 8396 | SORL1 | 6653 | PTPRN | 5798 | PAK3 | 5063 | NUP205 | 23165 | SUZ12 | 23512 | GNB4 | 59345 |
| TCAP | 8557 | SUPT6H | 6830 | RBM3 | 5935 | PCCB | 5096 | SCFD1 | 23256 | TAPT1 | 202018 | GOLGA2 | 2801 |
| B4GALT2 | 8704 | SVIL | 6840 | RPS3 | 6188 | PDK3 | 5165 | DDAH1 | 23576 | TASP1 | 55617 | GRPEL1 | 80273 |
| B3GALT2 | 8707 | SYK | 6850 | RPS5 | 6193 | PHF2 | 5253 | NUP62 | 23636 | UBR2 | 23304 | H1FX | 8971 |
| CRADD | 8738 | Tbcd | 6904 | RPS10 | 6204 | PKD1 | 5310 | PPP1R15A | 23645 | VPS33A | 65082 | H2AFJ | 55766 |
| HRK | 8739 | TCF20 | 6942 | RPS14 | 6208 | POLA1 | 5422 | MRPL22 | 29093 | WWP2 | 11060 | H2BFS | 54145 |
| RAB7L1 | 8934 | TFAP2A | 7020 | RPS16 | 6217 | PRKACA | 5566 | WAC | 51322 | YLPM1 | 56252 | HERC5 | 51191 |
| MPZL1 | 9019 | TFAP2C | 7022 | RPS24 | 6229 | PRKCD | 5580 | FZR1 | 51343 | ZFAND3 | 60685 | HIST1H1A | 3024 |
| SART1 | 9092 | tgfbr1 | 7046 | RPS27A | 6233 | PRKCI | 5584 | EIF3I | 51386 | ZFHX3 | 463 | HIST1H1T | 3010 |
| USP6 | 9098 | UBC | 7316 | SAFB | 6294 | MAPK1 | 5594 | ARID4B | 51742 | ZNF7 | 7553 | HIST1H2AC | 8334 |
| BUB3 | 9184 | UMPS | 7372 | SARS | 6301 | MAP2K2 | 5605 | SEC61A2 | 55176 | | | HIST1H2AD | 3013 |
| COPB2 | 9276 | WNT5A | 7474 | SCNN1D | 6339 | MAP2K3 | 5606 | VAC14 | 55697 | | | HIST1H2AE | 3012 |
| EFTUD2 | 9343 | Xbp1 | 7494 | SELPLG | 6404 | MAP2K5 | 5607 | PCID2 | 55795 | | | HIST1H2AG | 8969 |
| ONECUT2 | 9480 | ZIC1 | 7545 | SFTPB | 6439 | EIF2AK2 | 5610 | NXT2 | 55916 | | | HIST1H2AH | 85235 |
| APBA3 | 9546 | ZKSCAN1 | 7586 | SLC12A4 | 6560 | PSMD1 | 5707 | RPTOR | 57521 | | | HIST1H2AI | 8329 |
| ATP6V1G1 | 9550 | TRIM25 | 7706 | SNRP70 | 6625 | PTMA | 5757 | RFX7 | 64864 | | | HIST1H2AJ | 8331 |
| MTFR1 | 9650 | TRIM26 | 7726 | SNRPF | 6636 | PTPRM | 5797 | AAGAB | 79719 | | | HIST1H2AK | 8330 |
| PLCH2 | 9651 | ZNF217 | 7764 | SON | 6651 | PTPRN | 5798 | ZBTB3 | 79842 | | | HIST1H2AL | 8332 |
| ZNF432 | 9668 | VPS24 | 7844 | TRIM21 | 6737 | PTS | 5805 | BICC1 | 80114 | | | HIST1H2BB | 3018 |
| KIAA0391 | 9692 | HMGA2 | 8091 | SSTR5 | 6755 | PRPH2 | 5961 | RAB1B | 81876 | | | HIST1H2BC | 8347 |
| AQR | 9716 | C21orf33 | 8209 | SUPT6H | 6830 | RING1 | 6015 | ATPBD4 | 89978 | | | HIST1H2BD | 3017 |
| KIAA0652 | 9776 | stk24 | 8428 | TCF3 | 6929 | ROCK1 | 6093 | CCDC149 | 91050 | | | HIST1H2BD | 3017 |
| FAM38A | 9780 | PIK3R3 | 8503 | TFE3 | 7030 | RPS6KA2 | 6196 | C20orf54 | 113278 | | | HIST1H2BE | 8344 |
| SPCS2 | 9789 | bhlhe40 | 8553 | TK2 | 7084 | RPS10 | 6204 | OSBPL1A | 114876 | | | HIST1H2BF | 8343 |
| PHYHIP | 9796 | OASL | 8638 | WNT9A | 7483 | RPS20 | 6224 | HIST2H3C | 126961 | | | HIST1H2BH | 8345 |
| NUPL1 | 9818 | RIPK2 | 8767 | XPNPEP1 | 7511 | SCN3A | 6328 | C9orf85 | 138241 | | | HIST1H2BI | 8346 |
| IQSEC1 | 9922 | TNFRSF6B | 8771 | XPO1 | 7514 | SCN8A | 6334 | DCP2 | 167227 | | | HIST1H2BJ | 8970 |
| DCLRE1A | 9937 | Riok3 | 8780 | DAP3 | 7818 | SCNN1G | 6340 | KRTCAP2 | 200185 | | | HIST1H2BK | 85236 |
| TRIM28 | 10155 | TNFRSF10D | 8793 | BRPF1 | 7862 | CCL13 | 6357 | ANKS6 | 203286 | | | HIST1H2BL | 8340 |
| NUTF2 | 10204 | SYNJ1 | 8867 | LHX3 | 8022 | SGCA | 6442 | PHACTR1 | 221692 | | | HIST1H2BM | 8342 |
| RBM14 | 10432 | Eif2s2 | 8894 | ARD1A | 8260 | SGK1 | 6446 | C3orf23 | 285343 | | | HIST1H2BN | 8341 |
| OLFM1 | 10439 | WASF1 | 8936 | AXIN1 | 8312 | SIAH2 | 6478 | LIN9 | 286826 | | | HIST1H2BO | 8348 |
| NXF1 | 10482 | RPS6KA4 | 8986 | HIST1H2BN | 8341 | SMARCD3 | 6604 | BARHL2 | 343472 | | | HIST1H4A | 8359 |
| SLC35A1 | 10559 | OSMR | 9180 | NCK2 | 8440 | SUMO2 | 6613 | RSPH4A | 345895 | | | HIST1H4B | 8366 |
| PRPF8 | 10594 | Bub3 | 9184 | EIF3A | 8661 | FSCN1 | 6624 | NUP43 | 348995 | | | HIST1H4C | 8364 |
| SLC17A3 | 10786 | RPS6KA5 | 9252 | EIF3C | 8663 | SNRNP70 | 6625 | RPL13P12 | 388344 | | | HIST1H4D | 8360 |
| CPLX1 | 10815 | CYTH2 | 9266 | EIF3F | 8665 | SNRPA1 | 6627 | RPS9 | 388556 | | | HIST1H4E | 8367 |
| HPSE | 10855 | GPR56 | 9289 | EIF3G | 8666 | CDKL5 | 6792 | EIF3F | 390282 | | | HIST1H4F | 8361 |
| RAB10 | 10890 | Tceal1 | 9338 | SIGLEC5 | 8778 | STX5 | 6811 | HSPA5 | 400750 | | | HIST1H4H | 8365 |
| COPS6 | 10980 | FADS2 | 9415 | TNFRSF18 | 8784 | TEAD3 | 7005 | EIF2AK4 | 440275 | | | HIST1H4I | 8294 |
| SF3B2 | 10992 | eif2ak3 | 9451 | VNN2 | 8875 | TPT1 | 7178 | RPS7P3 | 440732 | | | HIST1H4J | 8363 |
| ERLIN2 | 11160 | Adamts1 | 9510 | NAE1 | 8883 | TXK | 7294 | | | | | HIST1H4L | 8368 |
| INMT | 11185 | MICAL2 | 9645 | BAIAP3 | 8938 | SUMO1 | 7341 | | | | | HIST2H2AA3 | 8337 |
| AKAP11 | 11215 | KIAA0406 | 9675 | ARTN | 9048 | VEGFB | 7423 | | | | | HIST2H2AA4 | 723790 |
| CBX2 | 12416 | RAPGEF2 | 9693 | ATP6V0D1 | 9114 | MAP3K12 | 7786 | | | | | HIST2H2AC | 8338 |
| ITGA11 | 22801 | KIAA0226 | 9711 | LARGE | 9215 | NUP214 | 8021 | | | | | HIST2H2BE | 8349 |

| | | | | | | | | | | | | | | |
|---|---|---|---|---|---|---|---|---|---|---|---|---|---|---|
| COPZ1 | 22818 | mtss1 | 9788 | BZRAP1 | 9256 | HIST3H3 | 8290 | | | | | HIST2H2BF | 440689 | |
| COPG | 22820 | Tlk1 | 9874 | COPB2 | 9276 | RAD54L | 8438 | | | | | HIST2H4A | 8370 | |
| RBM34 | 23029 | CLEC2B | 9976 | MED14 | 9282 | CDC42BPA | 8476 | | | | | HIST2H4B | 554313 | |
| MCF2L | 23263 | Snupn | 10073 | SLC22A6 | 9356 | CDK10 | 8558 | | | | | HIST3H2A | 92815 | |
| SF3B3 | 23450 | ptpru | 10076 | LONP1 | 9361 | KHSRP | 8570 | | | | | HIST3H2BB | 128312 | |
| SF3B1 | 23451 | Rasgrp2 | 10235 | SLC9A3R1 | 9368 | STX10 | 8677 | | | | | HLA-A | 3105 | |
| GCAT | 23464 | CALCOCO2 | 10241 | ISG15 | 9636 | SYNGAP1 | 8831 | | | | | HLA-DRA | 3122 | |
| ETHE1 | 23475 | SPRY1 | 10252 | CLSTN3 | 9746 | CFLAR | 8837 | | | | | HLA-DRA | 3122 | |
| RPL13A | 23521 | TRIM22 | 10346 | EIF4A3 | 9775 | ATP6V0D1 | 9114 | | | | | HLA-DRB1 | 3123 | |
| ADAT1 | 23536 | NOD1 | 10392 | HARBI1 | 9776 | RABEP1 | 9135 | | | | | HLA-DRB1 | 3123 | |
| BACE1 | 23621 | Tab1 | 10454 | TRIM14 | 9830 | DYRK1B | 9149 | | | | | HMBOX1 | 79618 | |
| IFIT5 | 24138 | NXF1 | 10482 | LPPR4 | 9890 | PCSK7 | 9159 | | | | | HNRNPA0 | 10949 | |
| SAMM50 | 25813 | RBCK1 | 10616 | FGF19 | 9965 | OSMR | 9180 | | | | | HNRNPA1 | 3178 | |
| ZNF473 | 25888 | IVNS1ABP | 10625 | MED6 | 10001 | DCLK1 | 9201 | | | | | HNRNPA2B1 | 3181 | |
| CNOT10 | 25904 | GMEB1 | 10691 | RANBP9 | 10048 | RAB11B | 9230 | | | | | HNRNPC | 3183 | |
| CHMP2B | 25978 | pold3 | 10714 | TRAP1 | 10131 | DLG5 | 9231 | | | | | HNRNPL | 3191 | |
| PRPF31 | 26121 | mthfd2 | 10797 | ATP6AP2 | 10159 | BZRAP1 | 9256 | | | | | HNRNPR | 10236 | |
| C14orf109 | 26175 | WASF3 | 10810 | TNK2 | 10188 | COPB2 | 9276 | | | | | HNRNPU | 3192 | |
| FBXO22 | 26263 | rab40b | 10966 | DHRS2 | 10202 | MAP4K4 | 9448 | | | | | HNRPA3 | 220988 | |
| SIGLEC9 | 27180 | KIF2C | 11004 | PSMD14 | 10213 | HAND2 | 9464 | | | | | HNRPAB | 3182 | |
| SULT1C4 | 27233 | Il24 | 11009 | SIGMAR1 | 10280 | ADAMTS2 | 9509 | | | | | HNRPD | 3184 | |
| GPKOW | 27238 | PRAF2 | 11230 | SF3A1 | 10291 | CLOCK | 9575 | | | | | HNRPDL | 9987 | |
| PDLIM3 | 27295 | stk38 | 11329 | KATNB1 | 10300 | CDC42BPB | 9578 | | | | | HNRPF | 3185 | |
| MOCS3 | 27304 | WDR37 | 22884 | BAIAP2 | 10458 | AATK | 9625 | | | | | HNRPH1 | 3187 | |
| RAB30 | 27314 | DAAM1 | 23002 | TIMM44 | 10469 | IKBKE | 9641 | | | | | HNRPH2 | 3188 | |
| ANGPTL3 | 27329 | lpin1 | 23175 | NXF1 | 10482 | OXSR1 | 9943 | | | | | HNRPH3 | 3189 | |
| NT5C | 30833 | Rftn1 | 23180 | FBXW10 | 10517 | NUP153 | 9972 | | | | | HNRPK | 3190 | |
| TRIM17 | 51127 | PHF3 | 23469 | PRPF8 | 10594 | CHAF1A | 10036 | | | | | HNRPUL1 | 11100 | |
| C5orf45 | 51149 | ZFYVE26 | 23503 | PAICS | 10606 | SAE1 | 10055 | | | | | HOOK1 | 51361 | |
| CYB5R4 | 51167 | RPL13AP7 | 23521 | TBL3 | 10607 | PPIF | 10105 | | | | | HRNR | 388697 | |
| FAM13B | 51306 | ifit5 | 24138 | CXCR6 | 10663 | HIPK3 | 10114 | | | | | HSD17B10 | 3028 | |
| CRNKL1 | 51340 | CCRN4L | 25819 | AHCYL1 | 10768 | TRIM28 | 10155 | | | | | HSP90AA1 | 3320 | |
| C14orf166 | 51367 | TMEM186 | 25880 | TXNL4A | 10907 | ATP6AP2 | 10159 | | | | | HSP90AB1 | 3326 | |
| VPS54 | 51542 | CLIC4 | 25932 | CENTA1 | 11033 | RBM5 | 10181 | | | | | HSPA1L | 3305 | |
| PIGT | 51604 | TBC1D10B | 26000 | NUDT4 | 11163 | TNK2 | 10188 | | | | | HSPA6 | 3310 | |
| GINS2 | 51659 | c14orf109 | 26175 | RPL35 | 11224 | OPRS1 | 10280 | | | | | HSPA8 | 3312 | |
| LRP1B | 53353 | FBXW2 | 26190 | COPE | 11316 | SF3A1 | 10291 | | | | | HSPD1 | 3329 | |
| CNTN5 | 53942 | alg6 | 29929 | FKBP9 | 11328 | APC2 | 10297 | | | | | IGF2BP1 | 10642 | |
| NDUFB11 | 54539 | cutC | 51076 | COPG | 22820 | TUBB3 | 10381 | | | | | IGF2BP2 | 10644 | |
| NDE1 | 54820 | copz2 | 51226 | SNW1 | 22938 | ERN2 | 10595 | | | | | IGF2BP3 | 10643 | |
| SLC41A3 | 54946 | TAOK3 | 51347 | MYT1L | 23040 | RBCK1 | 10616 | | | | | IKZF3 | 22806 | |
| CDK5RAP2 | 55755 | PCYOX1 | 51449 | NUP205 | 23165 | NFAT5 | 10725 | | | | | ILF2 | 3608 | |
| TBC1D23 | 55773 | Pcf11 | 51585 | STAB1 | 23166 | PLK4 | 10733 | | | | | ILF3 | 3609 | |
| MBNL3 | 55796 | NBAS | 51594 | KIAA0664 | 23277 | NEK6 | 10783 | | | | | IRS4 | 8471 | |
| PCDHGA1 | 56114 | lsr | 51599 | LARP1 | 23367 | CD3EAP | 10849 | | | | | ISG15 | 9636 | |
| PPAN | 56342 | inpp5k | 51763 | MLYCD | 23417 | CIT | 11113 | | | | | JTV1 | 7965 | |
| TMEM167B | 56900 | PION | 54103 | SF3B1 | 23451 | IRAK3 | 11213 | | | | | KCNRG | 283518 | |
| C11orf60 | 56912 | Fam35a | 54537 | ABCB10 | 23456 | AKAP13 | 11214 | | | | | KHDRBS1 | 10657 | |
| NUP107 | 57122 | Ing3 | 54556 | TRAM1 | 23471 | COPG | 22820 | | | | | KHDRBS3 | 10656 | |
| SCYL3 | 57147 | zcchc8 | 55596 | SRRM2 | 23524 | SMG1 | 23049 | | | | | KIAA1143 | 57456 | |
| WDR18 | 57418 | Polr3b | 55703 | PIK3R5 | 23533 | TBC1D1 | 23216 | | | | | KIAA1967 | 57805 | |
| ZNF492 | 57615 | KBTBD4 | 55709 | CHST5 | 23563 | UBR4 | 23352 | | | | | KPNA1 | 3836 | |
| CWC22 | 57703 | LIN37 | 55957 | IL17RA | 23765 | NUDCD3 | 23386 | | | | | KPNA3 | 3839 | |
| LSM2 | 57819 | ccnl1 | 57018 | FKBP8 | 23770 | KIAA0999 | 23387 | | | | | KPNA6 | 23633 | |
| C19orf29 | 58509 | PELI1 | 57162 | PART1 | 25859 | PIP5K1C | 23396 | | | | | KPNB1 | 3837 | |
| UBL5 | 59286 | Zmiz1 | 57178 | CNRIP1 | 25927 | TNPO3 | 23534 | | | | | LDHB | 3945 | |
| ZNF350 | 59348 | ZNF512B | 57473 | CLIC4 | 25932 | CCRK | 23552 | | | | | LENG8 | 114823 | |
| DCLRE1B | 64858 | KIAA1609 | 57707 | HERC4 | 26091 | DAPK2 | 23604 | | | | | LEO1 | 123169 | |
| FAM173A | 65990 | Scpep1 | 59342 | FBXW2 | 26190 | IL17RA | 23765 | | | | | LGALS3BP | 3959 | |
| FN3KRP | 79672 | Sav1 | 60485 | CNNM1 | 26507 | FKBP8 | 23770 | | | | | LNX2 | 222484 | |
| TBL1XR1 | 79718 | rtp4 | 64108 | ATP2C1 | 27032 | HECTD1 | 25831 | | | | | LYPLA1 | 10434 | |
| ZNF552 | 79818 | PLA2G2F | 64600 | TRIB2 | 28951 | CACNG4 | 27092 | | | | | LZTS2 | 84445 | |
| TMEM62 | 80021 | USP46 | 64854 | FHOD1 | 29109 | STK39 | 27347 | | | | | MAGEA11 | 4110 | |
| FHOD3 | 80206 | ARMCX5 | 64860 | RACGAP1 | 29127 | HIPK2 | 28996 | | | | | MAGEA12 | 4111 | |
| LY6G6C | 80740 | RSRC2 | 65117 | TBX21 | 30009 | C16orf72 | 29035 | | | | | MAGEA6 | 4105 | |
| STARD5 | 80765 | CHAC1 | 79094 | KCNIP3 | 30818 | TBK1 | 29110 | | | | | MAGED1 | 9500 | |
| PRRT1 | 80863 | hectd3 | 79654 | AIG1 | 51390 | RACGAP1 | 29127 | | | | | MAPK9 | 5601 | |
| GDPD5 | 81544 | ZMAT4 | 79698 | SNX9 | 51429 | ANAPC2 | 29882 | | | | | MARCKS | 4082 | |
| C6orf62 | 81688 | TNIP3 | 79931 | RAB6B | 51560 | NRBP1 | 29959 | | | | | MATR3 | 9782 | |
| RPS6KL1 | 83694 | ULBP1 | 80329 | SF3B14 | 51639 | HUNK | 30811 | | | | | MDH2 | 4191 | |
| FRMD8 | 83786 | STARD5 | 80765 | LRP1B | 53353 | ST6GALNAC6 | 30815 | | | | | MEOX2 | 4223 | |
| PCGF6 | 84108 | TLR10 | 81793 | NLE1 | 54475 | PIK3R4 | 30849 | | | | | MGC16075 | 84847 | |
| ZNF414 | 84330 | tcf7l1 | 83439 | FLJ11235 | 54508 | MINK1 | 50488 | | | | | MIF | 4282 | |
| ALG10 | 84920 | lzts2 | 84445 | PCDH18 | 54510 | TXNDC11 | 51061 | | | | | MIPOL1 | 145282 | |
| SELI | 85465 | IL1F10 | 84639 | APBB1IP | 54518 | NAGPA | 51172 | | | | | MLH1 | 4292 | |
| CCDC74A | 90557 | ABCC10 | 89845 | PPP1R14D | 54866 | 02-Mar | 51257 | | | | | MOV10 | 4343 | |
| PAGE5 | 90737 | IL33 | 90865 | SMU1 | 55234 | AIG1 | 51390 | | | | | MRPL12 | 6182 | |
| ESAM | 90952 | PLCD3 | 113026 | ITLN1 | 55600 | TRPV2 | 51393 | | | | | MTAP | 4507 | |
| CCDC74B | 91409 | AHNAK2 | 113146 | TRERF1 | 55809 | PRKAG2 | 51422 | | | | | MYH10 | 4628 | |
| YTHDC1 | 91746 | Apoa5 | 116519 | PSENEN | 55851 | C20orf111 | 51526 | | | | | MYH14 | 79784 | |
| MYOCD | 93649 | Il31ra | 133396 | DMAP1 | 55929 | ADAMTSL4 | 54507 | | | | | MYH9 | 4627 | |
| TGS1 | 96764 | LOC285830 | 285830 | SULF2 | 55959 | PPP1R12C | 54776 | | | | | MYL6 | 4637 | |
| BIRC8 | 112401 | NLRP10 | 338322 | NXF3 | 56000 | PPP1R14D | 54866 | | | | | MYO1C | 4641 | |
| C21orf82 | 114036 | NLRP14 | 338323 | RETN | 56729 | C2orf42 | 54980 | | | | | NCALD | 83988 | |
| CARD16 | 114769 | INCA1 | 388324 | C21orf7 | 56911 | C1orf159 | 54991 | | | | | NCAPH2 | 29781 | |
| STK11IP | 114790 | | | XAB2 | 56949 | PANK4 | 55229 | | | | | NDUFS3 | 4722 | |
| NOSTRIN | 115677 | | | RNF150 | 57484 | NAGK | 55577 | | | | | NEFH | 4744 | |
| ZNF653 | 115950 | | | HEG | 57493 | SLC48A1 | 55652 | | | | | NISCH | 11188 | |
| SFXN2 | 118980 | | | FAM135A | 57579 | USE1 | 55850 | | | | | NKRF | 55922 | |
| CCDC78 | 124093 | | | CRAMP1L | 57585 | PSENEN | 55851 | | | | | NOL5A | 10528 | |
| RAVER1 | 125950 | | | ZBTB2 | 57621 | PBK | 55872 | | | | | NPM1 | 4869 | |
| EVC2 | 132884 | | | SNX6 | 58533 | STK31 | 56164 | | | | | NRF1 | 4899 | |

| | | | | | | | | | | | |
|---|---|---|---|---|---|---|---|---|---|---|---|
| SAMD8 | 142891 | | | BACH2 | 60468 | IL1F9 | 56300 | | | PABPC1 | 26986 |
| C22orf15 | 150248 | | | AKTIP | 64400 | ANKRD7 | 56311 | | | PABPC4 | 8761 |
| NFXL1 | 152518 | | | ENGASE | 64772 | KCNK12 | 56660 | | | PABPN1 | 8106 |
| SLC36A2 | 153201 | | | P2RY12 | 64805 | UBQLN4 | 56893 | | | PAF1 | 54623 |
| AMOTL1 | 154810 | | | RBM42 | 79171 | CABC1 | 56997 | | | PAICS | 10606 |
| ATP6V0E2 | 155066 | | | LASS4 | 79603 | AGTRAP | 57085 | | | PARP1 | 142 |
| ZNF567 | 163081 | | | CCDC51 | 79714 | GOPC | 57120 | | | PCBP1 | 5093 |
| C1orf55 | 163859 | | | FLJ23554 | 79864 | WDR18 | 57418 | | | PCBP2 | 5094 |
| SLU7 | 193116 | | | C10orf57 | 80195 | NLGN4X | 57502 | | | PCBP3 | 54039 |
| MLKL | 197259 | | | LY6G6C | 80740 | MIB1 | 57534 | | | PCID2 | 55795 |
| SPAG17 | 200162 | | | AMN | 81693 | TAOK1 | 57551 | | | PFN1 | 5216 |
| KRTCAP2 | 200185 | | | FERMT3 | 83706 | FAM135A | 57579 | | | PGAM1 | 5223 |
| C2orf69 | 205327 | | | PRSS27 | 83886 | MID1IP1 | 58526 | | | PGAM2 | 5224 |
| C11orf82 | 220042 | | | TSSK6 | 83983 | RAB17 | 64284 | | | PGAM4 | 441531 |
| CADM2 | 239857 | | | USP42 | 84132 | VPS16 | 64601 | | | PGK1 | 5230 |
| STAC3 | 246329 | | | MORG1 | 84292 | NADK | 65220 | | | PIK3CA | 5290 |
| TCTEX1D2 | 255758 | | | ZNF512 | 84450 | EPS8L3 | 79574 | | | PIK3CB | 5291 |
| TPRX1 | 284355 | | | USMG5 | 84833 | ROGDI | 79641 | | | PIK3R1 | 5295 |
| CXorf59 | 286464 | | | LINGO1 | 84894 | LRRK1 | 79705 | | | PIK3R2 | 5296 |
| CALCOCO2 | 303479 | | | LRP11 | 84918 | CBLL1 | 79872 | | | PIK3R3 | 8503 |
| C1orf222 | 339457 | | | ATCAY | 85300 | CXorf21 | 80231 | | | PLAC8 | 51316 |
| LOC339524 | 339554 | | | PAQR8 | 85315 | ZNF436 | 80818 | | | PNMA1 | 9240 |
| FREM2 | 341640 | | | CCNB3 | 85417 | SGK196 | 84197 | | | POLR2A | 5430 |
| | | | | KCNK17 | 89822 | WDR34 | 89891 | | | PPIA | 5478 |
| | | | | NYD-SP25 | 89882 | FAM104B | 90736 | | | PPP1CA | 5499 |
| | | | | GGTLC2 | 91227 | G6PC3 | 92579 | | | PPP1CB | 5500 |
| | | | | DERL3 | 91319 | FBXO44 | 93611 | | | PPP1CC | 5501 |
| | | | | NEK9 | 91754 | ACRC | 93953 | | | PPP2R5C | 5527 |
| | | | | C19orf20 | 91978 | FOXQ1 | 94234 | | | PRDX6 | 9588 |
| | | | | DUSP27 | 92235 | LIMS3 | 96626 | | | PRKRA | 8575 |
| | | | | MEX3A | 92312 | DTX2 | 113878 | | | PRPF6 | 24148 |
| | | | | B3GNT7 | 93010 | PALM2 | 114299 | | | PTBP1 | 5725 |
| | | | | CYP2U1 | 113612 | CSMD3 | 114788 | | | PTMA | 5757 |
| | | | | GPR146 | 115330 | OSBPL6 | 114880 | | | QTRT1 | 81890 |
| | | | | FCHO2 | 115548 | PTPMT1 | 114971 | | | RABGEF1 | 27342 |
| | | | | C14orf172 | 115708 | ALPK2 | 115701 | | | RALY | 22913 |
| | | | | RFFL | 117584 | TOP1MT | 116447 | | | RAN | 5901 |
| | | | | AGAP4 | 119016 | GPR62 | 118442 | | | RARA | 5914 |
| | | | | SPATA2L | 124044 | C14orf28 | 122525 | | | RBMX | 27316 |
| | | | | ACVR1C | 130399 | CANT1 | 124583 | | | RBPMS | 11030 |
| | | | | C3orf31 | 132001 | OR10H4 | 126541 | | | RNF5 | 6048 |
| | | | | PASD1 | 139135 | UBXN10 | 127733 | | | RP11-78J21.1 | 144983 |
| | | | | C21orf121 | 150142 | C5orf38 | 153571 | | | RPA1 | 6117 |
| | | | | DCLK2 | 166614 | DCLK2 | 166614 | | | RPL11 | 6135 |
| | | | | TRIM60 | 166655 | PRSS35 | 167681 | | | RPL12 | 6136 |
| | | | | DTX3 | 196403 | TUBB | 203068 | | | RPL14 | 9045 |

| | | | | | | | | | | | | | | |
|---|---|---|---|---|---|---|---|---|---|---|---|---|---|---|
| | | | PAOX | 196743 | HIPK1 | 204851 | | | | | RPL15 | 6138 | | |
| | | | STXBP4 | 252983 | BRWD3 | 254065 | | | | | RPL17 | 6139 | | |
| | | | ANKK1 | 255239 | SYCE2 | 256126 | | | | | RPL18 | 6141 | | |
| | | | SPRYD4 | 283377 | KSR2 | 283455 | | | | | RPL19 | 6143 | | |
| | | | KIAA1267 | 284058 | RPL36AP49 | 284230 | | | | | RPL21 | 6144 | | |
| | | | NEK8 | 284086 | KLK9 | 284366 | | | | | RPL22 | 6146 | | |
| | | | UBAC2 | 337867 | DUPD1 | 338599 | | | | | RPL23 | 9349 | | |
| | | | MGC48998 | 339512 | SLC6A19 | 340024 | | | | | RPL23A | 6147 | | |
| | | | BARHL2 | 343472 | UNCX | 340260 | | | | | RPL24 | 6152 | | |
| | | | CA13 | 377677 | ENTPD8 | 377841 | | | | | RPL26 | 6154 | | |
| | | | LOC401431 | 401431 | SUMO4 | 387082 | | | | | RPL26L1 | 51121 | | |
| | | | FLJ25758 | 497049 | RP11-45B20.2 | 387911 | | | | | RPL27A | 6157 | | |
| | | | ZNF37B | 100129482 | NF1L2 | 401007 | | | | | RPL28 | 6158 | | |
| | | | VTRNA3P | 100144435 | OR51T1 | 401665 | | | | | RPL3 | 6122 | | |
| | | | | | LOC440396 | 440396 | | | | | RPL30 | 6156 | | |
| | | | | | MAP1LC3C | 440738 | | | | | RPL31 | 6160 | | |
| | | | | | LOC441239 | 441239 | | | | | RPL32 | 6161 | | |
| | | | | | OR4M1 | 441670 | | | | | RPL34 | 6164 | | |
| | | | | | ZNF862 | 643641 | | | | | RPL35 | 11224 | | |
| | | | | | LOC653712 | 653712 | | | | | RPL36A | 6173 | | |
| | | | | | LOC728683 | 728683 | | | | | RPL36AL | 6166 | | |
| | | | | | LOC730974 | 730974 | | | | | RPL37A | 6168 | | |
| | | | | | | | | | | | RPL38 | 6169 | | |
| | | | | | | | | | | | RPL4 | 6124 | | |
| | | | | | | | | | | | RPL6 | 6128 | | |
| | | | | | | | | | | | RPL7 | 6129 | | |
| | | | | | | | | | | | RPL7A | 6130 | | |
| | | | | | | | | | | | RPL8 | 6132 | | |
| | | | | | | | | | | | RPLP0 | 6175 | | |
| | | | | | | | | | | | RPLP0-like | 220717 | | |
| | | | | | | | | | | | RPLP1 | 6176 | | |
| | | | | | | | | | | | RPS10 | 6204 | | |
| | | | | | | | | | | | RPS11 | 6205 | | |
| | | | | | | | | | | | RPS12 | 6206 | | |
| | | | | | | | | | | | RPS16 | 6217 | | |
| | | | | | | | | | | | RPS18 | 6222 | | |
| | | | | | | | | | | | RPS19 | 6223 | | |
| | | | | | | | | | | | RPS2 | 6187 | | |
| | | | | | | | | | | | RPS20 | 6224 | | |
| | | | | | | | | | | | RPS21 | 6227 | | |
| | | | | | | | | | | | RPS24 | 6229 | | |
| | | | | | | | | | | | RPS26 | 6231 | | |
| | | | | | | | | | | | RPS27A | 6233 | | |
| | | | | | | | | | | | RPS3 | 6188 | | |
| | | | | | | | | | | | RPS4X | 6191 | | |
| | | | | | | | | | | | RPS5 | 6193 | | |

| Gene | ID |
|---|---|
| RPSA | 3921 |
| RTF1 | 23168 |
| RUVBL1 | 8607 |
| RUVBL2 | 10856 |
| SDCBP2 | 27111 |
| SECISBP2 | 79048 |
| SERBP1 | 26135 |
| SETBP1 | 26040 |
| SIAH1 | 6477 |
| SLC25A31 | 83447 |
| SLC25A4 | 291 |
| SLC25A5 | 292 |
| SLC25A6 | 293 |
| SMARCAL1 | 50485 |
| SNRPD3 | 6634 |
| SNRPF | 6636 |
| SP100 | 6672 |
| SSBP2 | 23635 |
| SSX2IP | 117178 |
| STAU1 | 6780 |
| STX5 | 6811 |
| SYNCRIP | 10492 |
| TACC1 | 6867 |
| TAF15 | 8148 |
| TARBP2 | 6895 |
| TCF12 | 6938 |
| TFCP2 | 7024 |
| TGFB1 | 7040 |
| THOC4 | 10189 |
| TIMM50 | 92609 |
| TMEM86B | 255043 |
| TPI1 | 7167 |
| TRAF1 | 7185 |
| TRAF2 | 7186 |
| TRIM13 | 10206 |
| TRIM25 | 7706 |
| TRIM27 | 5987 |
| TRIP6 | 7205 |
| TTC12 | 54970 |
| TUBA1A | 7846 |
| TUBA1B | 10376 |
| TUBA4A | 7277 |
| TUBB | 203068 |
| TUBB2C | 10383 |
| UBA52 | 7311 |
| UBB | 7314 |

| | | | | | | | | | | | | | |
|---|---|---|---|---|---|---|---|---|---|---|---|---|---|
| | | | | | | | | | | | UBC | 7316 | |
| | | | | | | | | | | | UBE2I | 7329 | |
| | | | | | | | | | | | UBE2L6 | 9246 | |
| | | | | | | | | | | | UPF1 | 5976 | |
| | | | | | | | | | | | UROS | 7390 | |
| | | | | | | | | | | | USHBP1 | 83878 | |
| | | | | | | | | | | | USP10 | 9100 | |
| | | | | | | | | | | | USP11 | 8237 | |
| | | | | | | | | | | | WDR18 | 57418 | |
| | | | | | | | | | | | WDR6 | 11180 | |
| | | | | | | | | | | | XPO1 | 7514 | |
| | | | | | | | | | | | XRN2 | 22803 | |
| | | | | | | | | | | | YBX1 | 4904 | |
| | | | | | | | | | | | YIPF6 | 286451 | |
| | | | | | | | | | | | YWHAB | 7529 | |
| | | | | | | | | | | | YWHAE | 7531 | |
| | | | | | | | | | | | YWHAG | 7532 | |
| | | | | | | | | | | | YWHAQ | 10971 | |
| | | | | | | | | | | | YWHAZ | 7534 | |
| | | | | | | | | | | | ZBTB25 | 7597 | |
| | | | | | | | | | | | ZMAT3 | 64393 | |
| | | | | | | | | | | | ZMAT4 | 79698 | |
| | | | | | | | | | | | ZNF346 | 23567 | |